\documentclass[final,5p,times,twocolumn,authoryear]{elsarticle}

\usepackage[authoryear]{natbib}
\usepackage{graphicx}
\usepackage{subcaption}
\usepackage{amssymb,amsmath}
\usepackage{pdflscape}
\usepackage{adjustbox}

\usepackage{booktabs}
\usepackage{lipsum}
\usepackage{epsfig}
\usepackage{txfonts}
\usepackage{xcolor}
\usepackage{tabularx}
\usepackage{footnote}
\usepackage{soul}
\usepackage{multirow}
\usepackage{multicol}
\usepackage{mathpazo}
\usepackage{lineno,hyperref}
\usepackage{textcomp}
\usepackage[utf8]{inputenc}
\usepackage[T1]{fontenc}
\usepackage{textgreek} 
\usepackage[utf8]{inputenc}
\usepackage{booktabs}

\usepackage{amsthm}

\journal{High Energy Astrophysics}

\begin{document}

\begin{frontmatter}



\title{Characterizing Short-Timescale Optical Variability in Non-blazar Active Galactic
Nucleus  PKS \,0521$-$36 Using TESS }


%

\author{Sikandar Akbar\corref{cor1}\fnref{label1}}
\ead{darprince46@gmail.com}
\affiliation[label1]{Department of Physics, University of Kashmir, Srinagar 190006, India}
\cortext[cor1]{Corresponding author: Sikandar Akbar}

\author{Zahir Shah\corref{cor1}\fnref{label2}}
\ead{shahzahir4@gmail.com}
\affiliation[label2]{Department of Physics, Central University of Kashmir, Ganderbal 191201, India.}

\cortext[cor1]{Corresponding author: Zahir Shah}



\begin{abstract}

We present a systematic analysis of high-cadence optical light curves of the non-blazar AGN PKS~0521$-$36 obtained with TESS across three sectors: Sectors~5 and~6 (Cycle~1, 30~min cadence) and Sector~32 (Cycle~3, 10~min cadence).
The source exhibits moderate variability with $F_\mathrm{var} \approx 0.69$--$1.19\%$, consistent with a mildly beamed jet. The PSD in all sectors is better described by a bending power-law than a simple power law, with high-frequency slopes $\alpha_1 \approx 2.1$--2.9, indicating red-noise dominated variability. The flux distributions in all three sectors require two-component models, with the double log-normal providing the best description, suggesting the presence of two distinct optical flux states associated with quiescent jet emission and episodic flaring activity.
A significant QPO at $P = 2.838 \pm 0.078$~d is detected in Sector~5 at $>99.99\%$ confidence in the LSP, independently confirmed by WWZ ($2.839 \pm 0.110$~d) and supported at the $3\sigma$ level by DRW analysis. The signal spans $\sim$9 cycles within the 26.1-day baseline and is absent in Sectors~6 and~32, indicating a transient feature. The PSD bending frequency in sector~5 ($\nu_b \approx 0.308$~d$^{-1}$; $\sim$3.2~d) is in close agreement with the QPO period, suggesting that both features share a common physical origin. We interpret the detected oscillation within the framework of magnetohydrodynamic kink instabilities developing in the relativistic jet of PKS~0521$-$36, supported by the previously reported helicoidal motion in its optical jet. The moderate Doppler factor of the source ($\delta \approx 5$--10) naturally accounts for the observed day-scale period within this framework. This detection, combined with the previously reported $\gamma$-ray QPOs on timescales of months to years, suggests that PKS~0521$-$36 harbors a complex hierarchy of quasi-periodic variability mechanisms spanning several orders of magnitude in timescale, and, to the best of our knowledge, provides the first indication for an optical QPO in a non-blazar AGN with a directly imaged helical jet structure.

\end{abstract}




\end{frontmatter}




\section{Introduction}
\label{introduction}

Active galactic nuclei (AGNs) represent some of the most energetic and persistently 
variable phenomena in the universe, deriving their extraordinary luminosity from 
accretion processes onto supermassive black holes (SMBHs) residing at galactic centers 
\citep{2017A&ARv..25....2P, 2016ARA&A..54..725M}. Many AGNs launch powerful relativistic 
jets perpendicular to the accretion disk, producing non-thermal continuum emission 
spanning the full electromagnetic spectrum from radio wavelengths up to very high-energy 
$\gamma$-rays \citep{2019ARA&A..57..467B, 2019Galax...7...20B}. A subclass of AGNs 
known as blazars is characterized by jets oriented at very small angles to the observer's 
line of sight, resulting in strongly Doppler-boosted emission and extreme flux variability 
\citep{1995PASP..107..803U}. Nevertheless, a broader population of jet-hosting AGNs 
exists beyond this blazar category, encompassing sources with intermediate viewing 
geometries and jet properties that do not fit neatly into standard classification schemes.

PKS~0521$-$36 is a prominent example of such a source, exhibiting a combination of 
properties that have resisted straightforward classification over decades of study. The 
source shows conspicuous broad emission lines in the optical and ultraviolet regimes, 
accompanied by a steep radio spectrum \citep{1986ApJ...302..296K, 1995A&A...303..730S}. 
Its classification history has undergone several revisions: initially identified as an 
N~galaxy, it was subsequently reclassified as a BL~Lac object 
\citep{2015MNRAS.450.3975D}, and later proposed to occupy an intermediate position 
between a broad-line radio galaxy and a steep-spectrum radio quasar (SSRQ) 
\citep{2015MNRAS.450.3975D}. The most recent \textit{Fermi}-LAT source catalog 
(4FGL; \citealt{2020ApJS..247...33A}) reflects this ambiguity, as the accumulated 
data still do not support a definitive AGN class assignment for this source.

Radio interferometric observations demonstrate that the jet morphology of PKS~0521$-$36 
bears a closer resemblance to misaligned AGNs than to classical blazars, and 
dedicated studies have established that its jet Doppler boosting is only moderate 
\citep{2015MNRAS.450.3975D, 2019A&A...627A.148A}. The multiwavelength jet of 
PKS~0521$-$36 has attracted considerable attention across a wide range of frequencies. 
Resolved jet emission has been documented in the optical, X-ray, and radio bands 
\citep{1979MNRAS.188..415D, 1991ApJ...369L..55M, 2002MNRAS.335..142B, 
2017ques.workE..16L}, positioning this source among the most studied extragalactic 
jets. Crucially, the jet components detected across optical, near-infrared, and 
submillimeter wavelengths are found to be spatially coincident with the radio jet axis, 
pointing to a well-collimated outflow structure \citep{1999ApJ...526..643S, 
2009A&A...501..907F, 2016A&A...586A..70L}. Additionally, \citet{2017MNRAS.470L.107J} 
identified an S-shaped morphology in the jet, which may reflect either the interaction 
of the jet with the surrounding interstellar medium at kiloparsec scales, or intrinsic 
jet dynamics such as precession or a helical magnetic field configuration.

In the high-energy domain, \textit{Fermi}-LAT observations have established 
PKS~0521$-$36 as a variable $\gamma$-ray emitter capable of producing rapid flaring 
episodes. The first such event, detected in 2010 June, exhibited a flux doubling 
timescale of approximately 12~hr \citep{2015MNRAS.450.3975D}, followed by a second 
rapid flare in 2012 October with a characteristic timescale of $\sim$6~hr 
\citep{2019A&A...627A.148A}. A third episode of short-timescale $\gamma$-ray activity, 
with variability on a scale of $\sim$7~hr, was subsequently identified during an 
outburst in 2019 May \citep{2021ApJ...919...58Z}. The recurrence of such fast flux 
changes implies that the $\gamma$-ray emission originates in a highly compact region, 
which is particularly noteworthy given the moderate beaming of the jet in this source.

Beyond these episodic flares, the long-term $\gamma$-ray light curve of PKS~0521$-$36 
harbors quasi-periodic oscillatory behavior. Analyzing approximately 5.8~years of 
\textit{Fermi}-LAT data bracketed by two major outbursts (MJD~56317--58447), 
\citet{2021ApJ...919...58Z} uncovered a QPO at a period of $\sim$1.1~yr ($\sim$400 
days) at a confidence level of approximately $5\sigma$, corroborated independently 
by the Lomb--Scargle periodogram, weighted wavelet Z-transform, REDFIT analysis, and 
Gaussian process modeling. Extending the temporal baseline to the full 15~years of 
available \textit{Fermi}-LAT data, \citet{2023arXiv231212623S} identified three 
distinct QPO signatures at periods of $\sim$268, $\sim$295, and $\sim$806~days, 
where the longest period appears to correspond to the third harmonic of the 
shortest-period oscillation. The physical mechanisms responsible for such QPOs in AGNs 
are still actively debated. Among the scenarios discussed in the literature are 
quasi-periodic modulation of the Doppler factor driven by jet precession 
\citep{2004ApJ...615L...5R}, orbital dynamics in a gravitationally bound binary SMBH 
system \citep{1980Natur.287..307B}, and disk-driven instabilities such as 
Kelvin--Helmholtz modes propagating through the inner accretion region 
\citep{2013MNRAS.434.3487A}.

Collectively, the unusual classification status, moderate jet beaming, structured jet 
morphology, and richly variable emission of PKS~0521$-$36 render it an especially 
valuable target for probing the physical connection between jet structure and 
variability in non-blazar AGNs. Whereas prior work has predominantly focused on 
$\gamma$-ray timescales of months to years, the optical band at high cadence remains 
comparatively unexplored for this source. The \textit{Transiting Exoplanet Survey 
Satellite} (TESS) fills this observational gap through its uninterrupted, 
high time-resolution photometric coverage, offering a powerful means to investigate 
variability on timescales of hours to days and to search for quasi-periodic signals 
in the optical regime.

In this work, we present an analysis of TESS observations of PKS~0521$-$36 aimed at 
characterizing its short-timescale optical variability. Our study encompasses a 
systematic search for quasi-periodic oscillations, quantification of the fractional 
variability amplitude, and statistical characterization of the flux distribution. 
By combining these complementary techniques, we aim to shed new light on the 
stochastic and periodic components of the emission variability in this mildly 
beamed AGN, and to connect the observed optical behavior to the broader picture of 
jet activity in PKS~0521$-$36.

\begin{figure*}
    \centering
    \begin{subfigure}{\textwidth}
        \centering
        \includegraphics[width=0.8\textwidth]{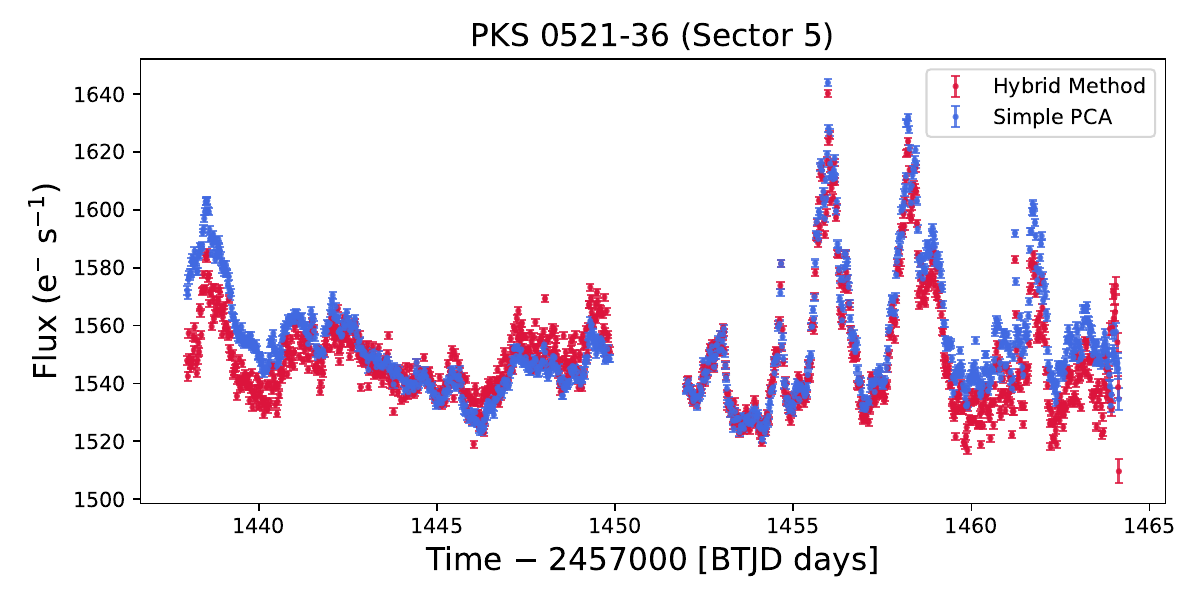}
        \caption{Sector~5 (1437.99--1464.14 BTJD; 
        15~November~2018 -- 11~December~2018; 30~min cadence)}
        \label{fig:lc_s5}
    \end{subfigure}
    
    \vspace{0.1cm}
    
    \begin{subfigure}{\textwidth}
        \centering
        \includegraphics[width=0.8\textwidth]{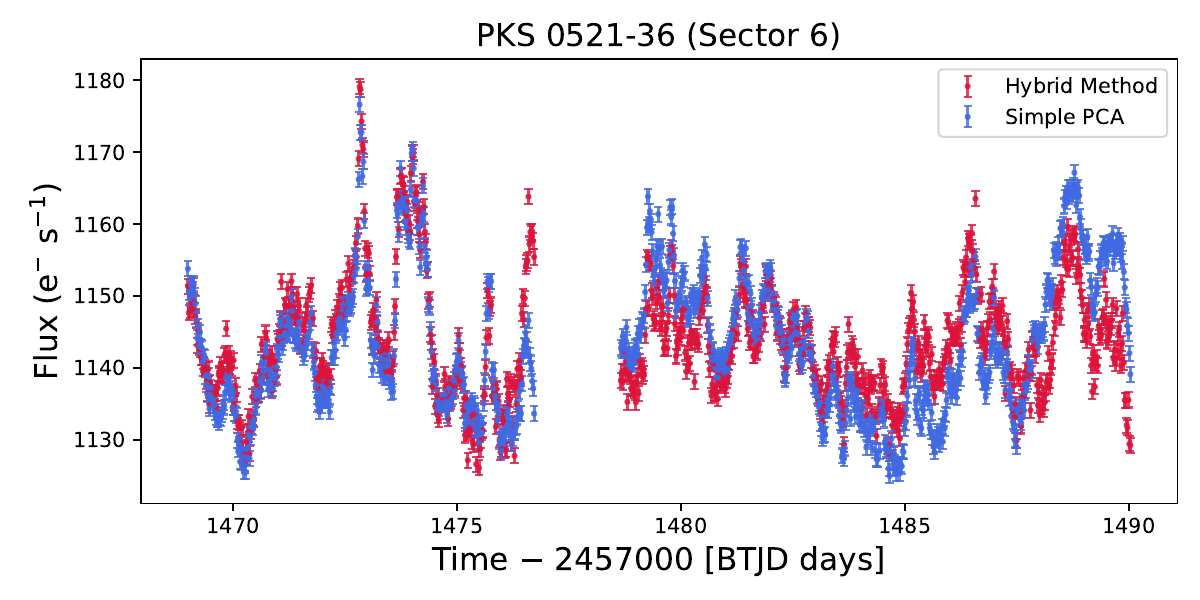}
        \caption{Sector~6 (1468.99--1490.03 BTJD; 
        12~December~2018 -- 6~January~2019; 30~min cadence)}
        \label{fig:lc_s6}
    \end{subfigure}
    
    \vspace{0.1cm}
    
    \begin{subfigure}{\textwidth}
        \centering
        \includegraphics[width=0.8\textwidth]{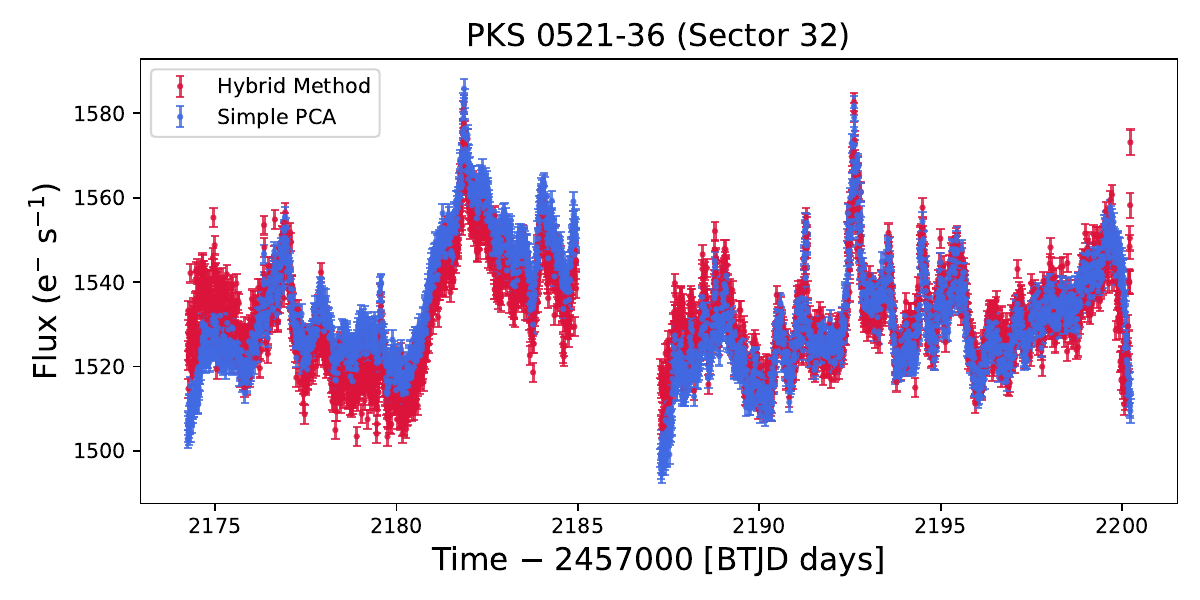}
        \caption{Sector~32 (2174.24--2200.23 BTJD; 
        19~November~2020 -- 16~December~2020; 10~min cadence)}
        \label{fig:lc_s32}
    \end{subfigure}
    
    \caption{Systematics-corrected TESS optical light curves of 
    PKS~0521$-$36 obtained using the \textsc{Quaver} pipeline for 
    Sectors~5, 6, and~32. The red points correspond to the full 
    hybrid reduction adopted for the timing analysis in this work, 
    while the blue points show the simple PCA (SPO) light curve 
    plotted for comparison. The time axis is given in BTJD 
    ($\mathrm{BTJD} = \mathrm{BJD} - 2457000$), and the flux is 
    in instrumental units of $\mathrm{e^{-}\,s^{-1}}$. The 
    prominent gap visible near the midpoint of each sector arises 
    from routine spacecraft data downlink operations. Both 
    reductions recover consistent variability patterns, supporting 
    the reliability of the full hybrid light curves used in the 
    subsequent analysis.}
    \label{fig:tess_lc}
\end{figure*}

\section{Observations and Data Reduction}
\label{sec:1}

\subsection{TESS Observations}
\label{sec:obs}

The optical photometric data analyzed in this study were acquired by the 
\textit{Transiting Exoplanet Survey Satellite} (TESS; \citealt{2015JATIS...1a4003R}), 
a space-borne observatory optimized for continuous, high-precision time-series 
photometry with near all-sky coverage. The instrument payload consists of four 
wide-field CCD cameras, each subtending $24^{\circ}\times24^{\circ}$ on the sky, 
which together yield a simultaneous field of view of $24^{\circ}\times96^{\circ}$. 
The full sky is divided into 26 partially overlapping sectors, distributed equally 
between the two hemispheres, each receiving roughly 27~days of nearly uninterrupted 
observation. The temporal sampling of the light curves has evolved across mission 
cycles: during the initial phase (2018--2019) data were recorded at 30~min cadence, 
whereas subsequent cycles (2020--present) introduced finer sampling at either 10~min 
or 2~min intervals, greatly enhancing the sensitivity to rapid flux variations.

A fundamental advantage of TESS over ground-based optical facilities is the absence 
of observational gaps introduced by the day--night cycle or seasonal visibility 
constraints, yielding quasi-continuous light curves with uniform temporal sampling. 
The total uninterrupted baseline achievable for a given target is governed primarily 
by its ecliptic latitude: sources located near the ecliptic plane are typically 
covered for a single sector duration, while those situated within the continuous 
viewing zones near the ecliptic poles can accumulate nearly a full year of 
consecutive monitoring. As the spacecraft alternates between the northern and southern 
hemispheres on an approximately annual basis, a subset of sources benefits from 
observations in multiple sectors separated by intervals of roughly one year. 
Regarding photometric sensitivity, the achievable precision scales with source 
brightness; for targets of comparable magnitude to PKS~0521$-$36, TESS is capable 
of reliably detecting variability at the $\sim$1--10\% level.

PKS~0521$-$36 was observed by TESS in three sectors spanning two 
separate cycles. Sector~5 and Sector~6 were obtained during Cycle~1, 
covering the intervals MJD~58437--58462 (15~November~2018 to 
11~December~2018) and MJD~58463--58489 (12~December~2018 to 
6~January~2019), respectively, both at a cadence of 30~min. 
Sector~32 was obtained during Cycle~3, covering the interval 
MJD~59172--59198 (19~November~2020 to 16~December~2020) at a 
cadence of 10~min. The target coordinates are 
$\alpha_{\rm J2000} = 80.7416^{\circ}$ and 
$\delta_{\rm J2000} = -36.4586^{\circ}$, corresponding to the TESS 
Input Catalog identifier TIC~167744793, as returned by the 
\textsc{Quaver} extraction query. In addition to these three sectors, 
PKS~0521$-$36 is also available in Sector~33 and Sector~98, but 
the present analysis is restricted to Sectors~5, 6, and~32, which 
together constitute the dataset used for the variability 
characterization and quasi-periodic oscillation search presented 
in this work.

\subsection{Data Reduction}
\label{sec:data_red}

The optical photometric data for PKS~0521$-$36 were extracted from TESS full-frame 
images (FFIs) using the open-source \textsc{Quaver} 
pipeline\footnote{\url{https://github.com/kristalynnesmith/quaver}} 
\citep{Smith_2023}, which is specifically developed for producing 
systematics-corrected light curves of AGNs from TESS observations. Standard TESS 
pipelines are primarily optimized for detecting periodic transit signals from 
exoplanets and are therefore not well-suited for the stochastic, aperiodic 
variability characteristic of AGNs \citep{2026MNRAS.545f1920T}. \textsc{Quaver} addresses 
this limitation by implementing a dedicated framework that avoids over- or 
under-fitting of AGN variability. Rather than requiring the full FFI data products, 
\textsc{Quaver} interfaces with the \textsc{TESSCut} package 
\citep{2019ASPC..523..397B} to retrieve a compact postage-stamp image cutout centered on 
the target coordinates, which substantially reduces the data volume. This 
cutout-based approach also permits the extraction aperture to be customized around 
the source, thereby limiting contamination from nearby field objects.

Once the cutout is retrieved, \textsc{Quaver} implements several tasks provided by 
the \textsc{Lightkurve} package \citep{2018soft12013L}. Central to the pipeline is 
a principal component analysis (PCA) framework for characterizing and removing 
instrumental systematics. PCA is applied to decompose pixel-level flux variations 
within the cutout into components associated with spacecraft systematics, scattered 
background light, and flux contributions from neighboring objects. A regression 
design matrix is constructed in which faint background pixels are used to track 
additive systematics, while the treatment of brighter pixels — which may carry 
astrophysical contamination or multiplicative trends — is handled separately. Before 
the correction is applied, \textsc{Quaver} guides the user through an interactive 
aperture selection step, in which Digital Sky Survey (DSS) contours are overlaid on 
the cutout to assist in defining an extraction region that minimizes blending and 
background leakage. The corrected light curve is then produced using 
\textsc{Lightkurve}'s \texttt{RegressionCorrector} \citep{Smith_2023}.

\textsc{Quaver} offers three distinct reduction modes: a simple PCA method, a simple 
hybrid method, and a full hybrid method \citep{Smith_2023}. 
The simple PCA method applies a user-specified number of principal components but 
cannot correct for most instrumental systematics and tends to over-fit long-term 
variability. The simple hybrid method accounts for background additive effects and 
instrumental systematics while preserving long-term variability trends, and has been 
shown to agree well with simultaneous ground-based observations 
\citep{Smith_2023}. The full hybrid method treats all systematics more 
rigorously and removes them directly from the source flux. In particular, it 
eliminates the effect of electronic crosstalk noise and other systematics that are 
significant in the high-frequency regime, making it well-suited for investigating 
rapid variability and quasi-periodic features \citep{Smith_2023}. Since the primary 
objective of this work is to characterize short-timescale variability and search for 
quasi-periodic oscillations -- tasks that require suppression of instrumental trends 
while preserving intrinsic rapid flux variations -- we therefore adopted the full hybrid 
reduction for all sectors analyzed here, following the approach of 
\citet{2026MNRAS.545f1920T}.

The target was queried by its common name (\texttt{pks0521-36}), and 
\textsc{Quaver} identified five available data products for PKS~0521$-$36 spanning 
TESS Sectors 5, 6, 32, 33, and 98. For the present study, we utilize light curves 
from three sectors: Sector~5 (15 November -- 11 December 2018) and Sector~6 (12 
December 2018 -- 6 January 2019), both observed at 30~min cadence during Cycle~1, 
and Sector~32 (19 November -- 16 December 2020), observed at 10~min cadence during 
Cycle~3. Each sector provides approximately 27~days of nearly continuous photometric 
coverage. However, each sector contains an inevitable gap, typically ranging from 
1 to 5~days, occurring when the spacecraft reorients toward the Earth to downlink 
the stored data \citep{2026MNRAS.545f1920T}. 
Diagnostic outputs from the pipeline — including the aperture selection overlay, 
regression components, and correction performance plots — were inspected for each 
sector to confirm that the dominant instrumental signatures were successfully removed 
without introducing spurious features into the light curve. As a consistency check, 
the full hybrid light curves were compared with those produced by the simple hybrid 
mode; both reductions recover consistent variability patterns across all three sectors, 
supporting the reliability of the corrected data used in the subsequent analysis. The resulting systematics-corrected light curves for 
Sectors~5, 6, and~32 are displayed in 
Figure~\ref{fig:tess_lc}, where both the full hybrid 
and simple PCA reductions are shown for comparison. 
All three sectors exhibit clear flux variability, which 
is examined quantitatively in 
Section~\ref{sec:analysis}.

\begin{table}
    \centering
    \caption{Fractional variability amplitude computed 
    sector-wise for PKS~0521$-$36 using the full hybrid 
    \textsc{Quaver} light curves.}
    \label{tab:fvar}
    \begin{tabular}{lccc}
    \hline
    Sector & Cadence & $F_{\rm var}$ (\%) & 
    $\Delta F_{\rm var}$ (\%) \\
    \hline
    Sector~5  & 30~min & $1.19$ & $0.002$ \\
    Sector~6  & 30~min & $0.69$ & $0.003$ \\
    Sector~32 & 10~min & $0.80$ & $0.003$ \\
    \hline
    \end{tabular}
\end{table}

\section{Data Analysis Methods}
\label{sec:analysis}

In this section, we describe the time-series analysis methods applied 
to the TESS light curves of PKS~0521$-$36. The analysis encompasses 
three complementary approaches. We first quantify the amplitude of 
optical flux variability through the fractional variability estimator 
$F_{\rm var}$. We then conduct a systematic search for 
QPOs using two independent periodicity detection 
techniques: the  Lomb--Scargle periodogram (LSP) and the 
weighted wavelet Z-transform (WWZ).  The results obtained by applying these methods 
to each sector individually are presented  in 
Section~\ref{sec:results}.
\subsection{Fractional variability}
To quantify the amplitude of the intrinsic 
variability in the TESS light curve of PKS~0521$-$36, we employ the fractional variability amplitude, $F_{\rm var}$, 
which is a widely used statistical estimator in AGN variability studies 
\citep{vaughan2003characterizing}.

For a light curve comprising $N$ flux measurements $x_i$, the total 
observed variance is expressed as
\begin{equation}
    S^2 = \frac{1}{N-1} \sum_{i=1}^{N} \left( x_i - \bar{x} \right)^2,
\end{equation}
where $\bar{x}$ denotes the mean flux. In practice, AGN light curves 
are subject to measurement uncertainties $\sigma_{{\rm err},i}$ that 
arise from instrumental noise, and these contribute additional scatter 
beyond the source's intrinsic variability. To isolate the genuine 
variability component, it is necessary to subtract this noise 
contribution. The excess variance $\sigma_{\rm XS}^2$, which represents 
the intrinsic source variance after removing the noise-induced component, 
is defined as \citep{nandra1997asca, edelson2002x}
\begin{equation}
    \sigma_{\rm XS}^2 = S^2 - \overline{\sigma_{\rm err}^2},
\end{equation}
where $\overline{\sigma_{\rm err}^2}$ is the mean squared measurement 
uncertainty, computed as
\begin{equation}
    \overline{\sigma_{\rm err}^2} = \frac{1}{N} \sum_{i=1}^{N} 
    \sigma_{{\rm err},i}^2.
\end{equation}
The fractional variability amplitude is then defined as the square root 
of the normalized excess variance $\sigma_{\rm NXS}^2 = 
\sigma_{\rm XS}^2 / \bar{x}^2$, yielding \citep{vaughan2003characterizing}
\begin{equation}
    F_{\rm var} = \sqrt{\frac{S^2 - \overline{\sigma_{\rm err}^2}}
    {\bar{x}^2}}.
\end{equation}
This quantity expresses the level of intrinsic variability as a fraction 
of the mean flux and is a linear statistic that facilitates comparison 
across different sources and observing epochs.

The uncertainty on the normalized excess variance is 
\citep{vaughan2003characterizing}
\begin{equation}
    {\rm err}\left(\sigma_{\rm NXS}^2\right) = \sqrt{
    \left( \sqrt{\frac{2}{N}} \cdot 
    \frac{\overline{\sigma_{\rm err}^2}}{\bar{x}^2} \right)^2
    + \left( \sqrt{\frac{\overline{\sigma_{\rm err}^2}}{N}} \cdot 
    \frac{2 F_{\rm var}}{\bar{x}} \right)^2},
\end{equation}
and the corresponding uncertainty on $F_{\rm var}$ is propagated 
following \citet{poutanen2008superorbital} and 
\citet{2018Galax...6....2B} as
\begin{equation}
    \Delta F_{\rm var} = \sqrt{F_{\rm var}^2 + 
    {\rm err}\left(\sigma_{\rm NXS}^2\right)} - F_{\rm var}.
\end{equation}
The $F_{\rm var}$ values computed sector-wise for PKS~0521$-$36 are 
presented in Table~\ref{tab:fvar}. The source exhibits a fractional 
variability amplitude ranging from $\sim$0.7\% to $\sim$1.2\% across 
the three sectors, indicating that PKS~0521$-$36 is moderately variable 
in the optical band on timescales of days during the epochs covered by 
the TESS observations.

\begin{figure*}
    \centering
    \begin{subfigure}{\textwidth}
        \centering
        \includegraphics[width=0.65\textwidth]{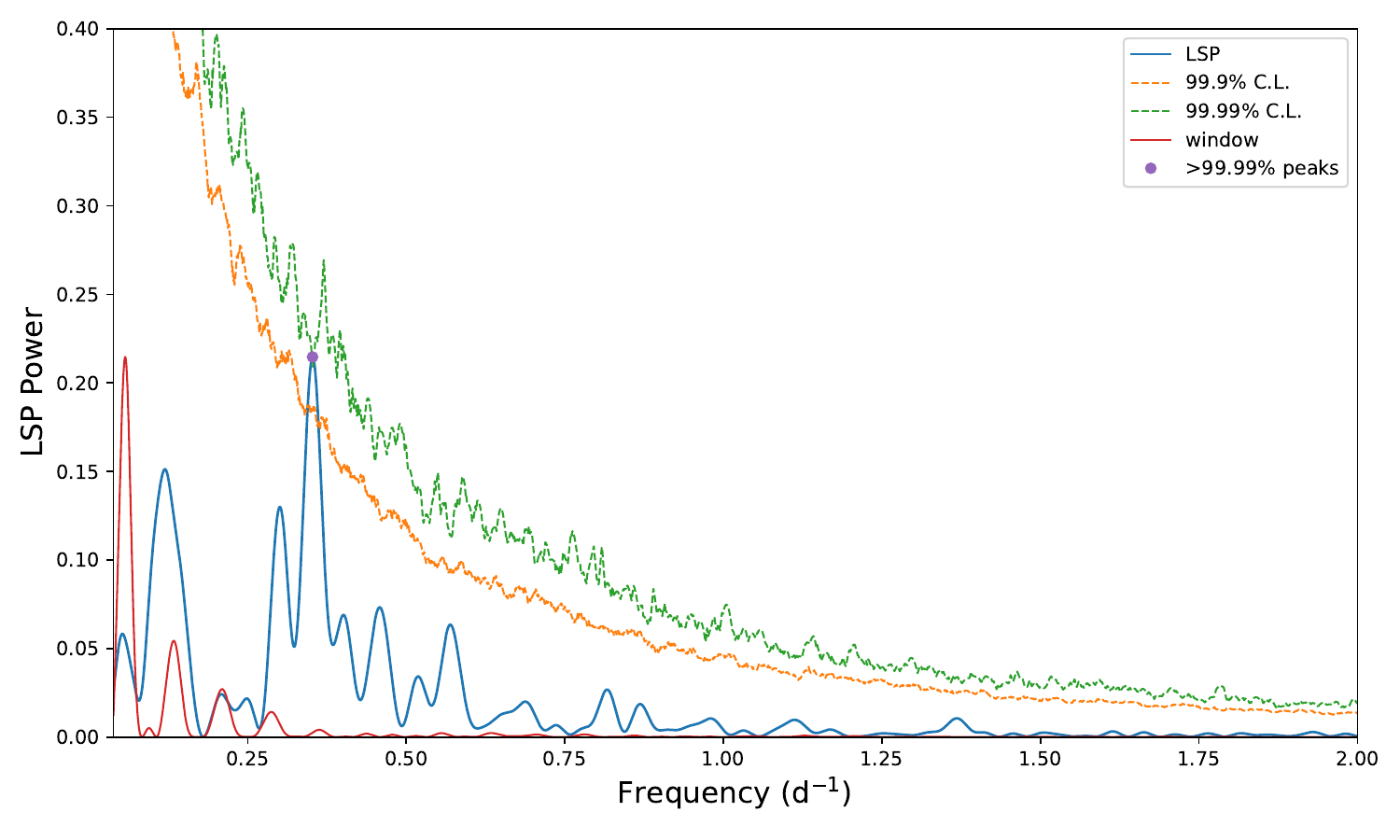}
        \caption{Sector~5 (1437.99--1464.14 BTJD; 30~min cadence). 
        The LSP shows a dominant peak exceeding the 99.99\% 
        confidence level (marked by the filled purple circle) 
        at $f \approx 0.35$~d$^{-1}$, corresponding to a period 
        of $\sim$2.84~d. The red solid curve shows the spectral 
        window function.}
        \label{fig:lsp_s5}
    \end{subfigure}
    
    \vspace{0.1cm}
    
    \begin{subfigure}{\textwidth}
        \centering
        \includegraphics[width=0.65\textwidth]{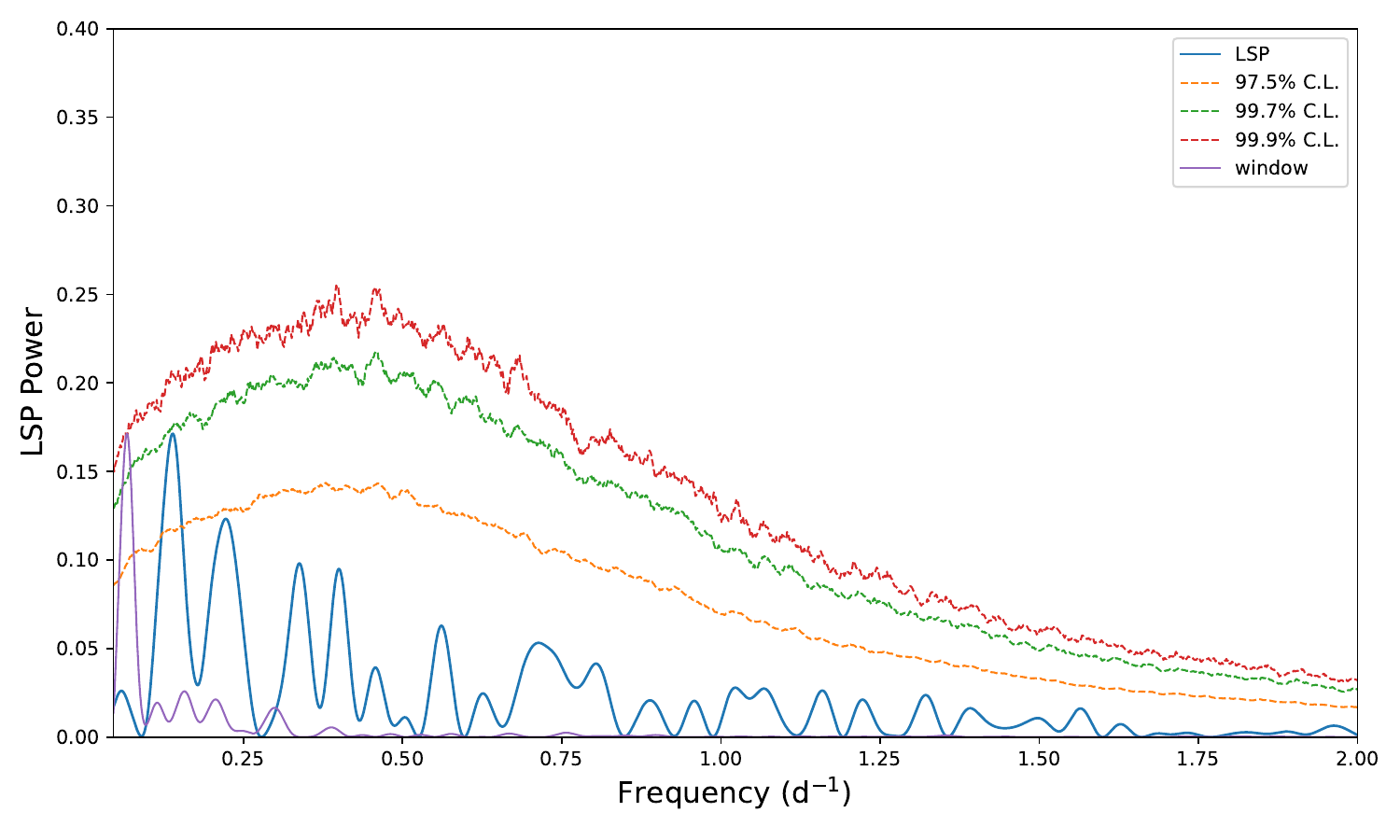}
        \caption{Sector~6 (1468.99--1490.03 BTJD; 30~min cadence). 
        No LSP peak exceeds any of the displayed significance 
        thresholds (97.5\%, 99.7\%, and 99.9\% confidence levels). 
        The purple solid curve shows the spectral window function.}
        \label{fig:lsp_s6}
    \end{subfigure}
    
    \vspace{0.1cm}
    
    \begin{subfigure}{\textwidth}
        \centering
        \includegraphics[width=0.65\textwidth]{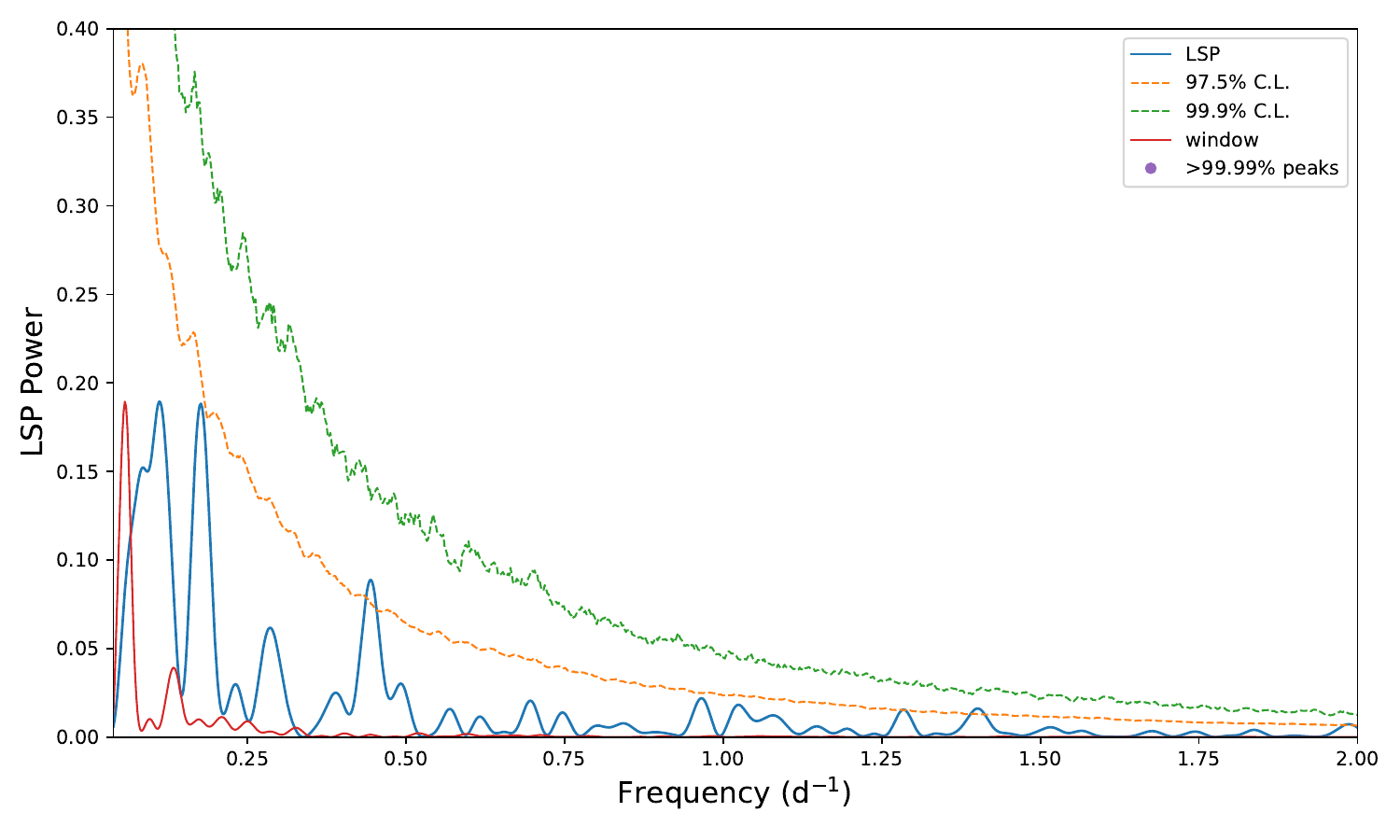}
        \caption{Sector~32 (2174.24--2200.23 BTJD; 10~min cadence). 
        No LSP peak exceeds the displayed significance thresholds 
        (97.5\% and 99.9\% confidence levels). The red solid curve 
        shows the spectral window function.}
        \label{fig:lsp_s32}
    \end{subfigure}
    
    \caption{Generalized Lomb--Scargle periodograms (LSP; blue 
    solid curves) of the \textsc{Quaver} full hybrid TESS light 
    curves of PKS~0521$-$36 for Sectors~5, 6, and~32. The dashed 
    curves show the Monte Carlo significance levels derived from 
    20\,000 simulated light curves following 
    \citet{2013MNRAS.433..907E}, with the specific 
    confidence levels indicated in each panel legend. The spectral window function is shown as a red solid curve in Sectors~5 and 32, and as a purple solid curve in Sector~6, which traces the power introduced by the 
    uneven sampling pattern of the TESS observations. Filled 
    purple circles mark peaks exceeding the 99.99\% confidence 
    level where present. A significant QPO candidate is detected 
    in Sector~5, while no statistically significant periodic 
    feature is identified in Sectors~6 or~32. The results are 
    discussed in Section~\ref{sec:results}.}
    \label{fig:lsp_all}
\end{figure*}

\subsection{Quasi-Periodic Oscillation Search}
\label{sec:qpo}

To search for quasi-periodic signatures in the TESS light curves 
of PKS~0521$-$36, we applied two complementary periodicity 
detection techniques: the generalized Lomb--Scargle periodogram 
(LSP) and the weighted wavelet Z-transform (WWZ).  The 
methodologies are described in the subsections below, and the 
corresponding results for each sector are presented in 
Section~\ref{sec:results}.

\subsubsection{Lomb--Scargle Periodogram}
\label{sec:lsp}

The Lomb--Scargle periodogram \citep{1976Ap&SS..39..447L, 
1982ApJ...263..835S} is one of the most widely employed techniques 
for detecting periodic signals in unevenly sampled time-series 
data. Its ability to handle irregular sampling makes it 
particularly well suited to space-based photometric observations 
such as those from TESS, which contain inherent data gaps. In 
this work, we used the \textsc{Astropy} implementation of the 
generalized Lomb--Scargle 
algorithm\footnote{\url{https://docs.astropy.org/en/stable/timeseries/lombscargle.html}}, 
incorporating the measured flux uncertainties into the 
computation to improve the reliability of the resulting 
periodograms. A detailed description of the underlying 
mathematical formalism is provided in \citet{2018ApJS..236...16V}.

Our application of the LSP follows the approach adopted in our earlier variability studies \citep{Nazir_2026,2026arXiv260120471A,zxgv-fzv5}. The frequency grid was constructed over the range 
$f_{\rm min} = 1/T$ to $f_{\rm max} = 1/(2\Delta T)$, where 
$T$ is the total temporal baseline of the light curve and 
$\Delta T$ is the characteristic sampling interval. The 
statistical significance of any peaks identified in the 
periodogram was assessed through the false-alarm probability 
(FAP), computed using the \texttt{LombScargle.false\_alarm\_probability()} 
routine from the \texttt{astropy.timeseries} module with 
\texttt{method=``baluev''}. This approach provides an analytic 
FAP estimate based on the extreme-value statistics formalism of 
\citet{2008MNRAS.385.1279B}, which accounts for the number of 
independent frequencies sampled across the periodogram. A 
periodic feature is considered statistically significant when 
its FAP falls below $10^{-3}$, corresponding to a confidence 
level exceeding 99.9\%. The period uncertainty for any 
significant peak is estimated by fitting a Gaussian profile 
to the peak and adopting the half-width at half-maximum (HWHM) 
as the error measure. The LSP results obtained for each sector 
are discussed in Section~\ref{sec:results}.

\subsubsection{Weighted Wavelet Z-Transform (WWZ)}
\label{sec:wwz}

The weighted wavelet $Z$-transform \citep[WWZ;][]{foster1996wavelets} 
provides a time--frequency representation of an unevenly sampled 
light curve by convolving the data with a localized oscillatory 
kernel. Unlike the LSP, which integrates over the full observational 
baseline, the WWZ is particularly well suited for detecting 
quasi-periodic signals whose amplitude varies with time, as it 
simultaneously constrains the characteristic timescale and the 
temporal interval over which the modulation is present. A genuine 
periodic component is expected to produce a localized enhancement 
in WWZ power that evolves as the signal strengthens or weakens 
over the course of the observation.

In this work, we adopted the abbreviated Morlet kernel,
\begin{equation}
f[\omega(t-\tau)] = \exp\!\left[i\omega(t-\tau)
-c\,\omega^{2}(t-\tau)^{2}\right],
\end{equation}
and computed the corresponding WWZ projection
\begin{equation}
W[\omega,\tau:x(t)] = \omega^{1/2}
\int x(t)\,f^{*}[\omega(t-\tau)]\,dt,
\end{equation}
where $f^{*}$ denotes the complex conjugate of the kernel, 
$\omega$ is the angular frequency, and $\tau$ represents the 
time offset. The analysis was carried out using the publicly 
available Python implementation of the WWZ 
algorithm\footnote{\url{https://github.com/eaydin/WWZ}}. 
Integrating the resulting two-dimensional WWZ power map along 
the time axis yields the time-averaged WWZ power spectrum, 
which provides an independent frequency-domain check 
complementary to the LSP. The peak frequency in the 
time-averaged spectrum was determined by fitting a Gaussian 
profile, with the half-width at half-maximum adopted as the 
frequency uncertainty. The WWZ results obtained for each sector 
are presented in Section~\ref{sec:results}.

\begin{figure*}
\centering
\includegraphics[width=0.9\textwidth]{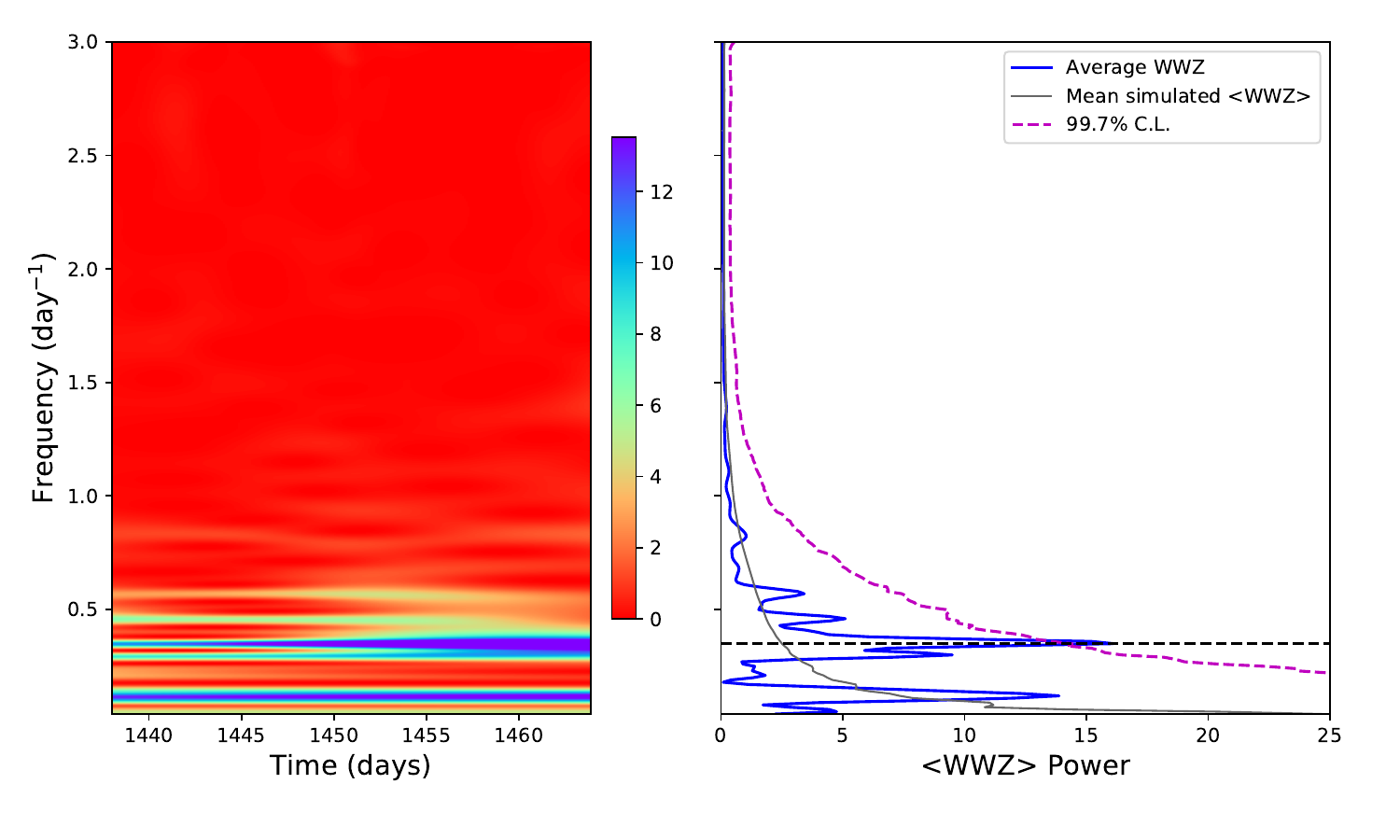}
\caption{Weighted wavelet Z-transform (WWZ) analysis of the 
4~hr binned TESS full hybrid light curve of PKS~0521$-$36 
for Sector~5. \textit{Left panel}: Two-dimensional WWZ 
time--frequency map showing the evolution of wavelet power 
as a function of time (BTJD) and frequency (d$^{-1}$). 
A persistent, localized band of enhanced power is visible 
near $f \approx 0.35$~d$^{-1}$, extending across the 
majority of the observing baseline, indicative of a 
quasi-periodic modulation sustained throughout the sector. 
\textit{Right panel}: Time-averaged WWZ power spectrum 
with the 99.7\% ($\sim$ 3$\sigma$) confidence level derived 
from $15\,000$ Monte Carlo simulations following 
\citet{2013MNRAS.433..907E}, shown as the 
dashed magenta curve.}
\label{fig:wwz_s5}
\end{figure*}

\subsection{Significance Evaluation}
\label{sec:sig_ev}

The optical light curves of AGNs are known to exhibit red-noise 
variability arising from stochastic processes in the jet or 
accretion flow, which can be characterized by a power-law power 
spectral density (PSD) of the form $P(\nu) \propto A\nu^{-\beta}$, 
where $\nu$ denotes the temporal frequency and $\beta > 0$ is the 
spectral index. In the presence of such 
red-noise backgrounds, spurious peaks can appear in periodograms 
at high significance levels purely by chance, making a careful 
statistical assessment of any candidate QPO feature essential 
before drawing physical conclusions.

\subsubsection{Monte Carlo Simulations}
\label{sec:mc}

To quantify the statistical significance of any peaks detected 
in the LSP and WWZ analyses, we employed a Monte Carlo simulation 
approach following the prescription of 
\citet{2013MNRAS.433..907E}. This method generates 
synthetic light curves that simultaneously reproduce both the 
power spectral density and the probability density function of 
the observed data, thereby preserving the stochastic character 
of the red-noise background. For the LSP analysis, a total of 
20\,000 synthetic light curves were generated, while 15\,000 
realizations were used for the WWZ significance assessment. 
The local significance of any candidate periodic feature was 
then estimated from the distribution of spectral powers at the 
corresponding frequency across the full ensemble of simulated 
light curves, yielding confidence levels at each trial frequency.

Since the LSP analysis of Sectors~6 and~32 does not reveal any 
peak exceeding the adopted significance threshold, the WWZ 
analysis and the DRW-based significance framework described 
below are applied only to Sector~5, where a candidate QPO 
feature is identified. The results of the Monte Carlo 
significance assessment for each sector are presented in 
Section~\ref{sec:results}.

\begin{figure*}
    \centering
    \includegraphics[width=0.9\textwidth]{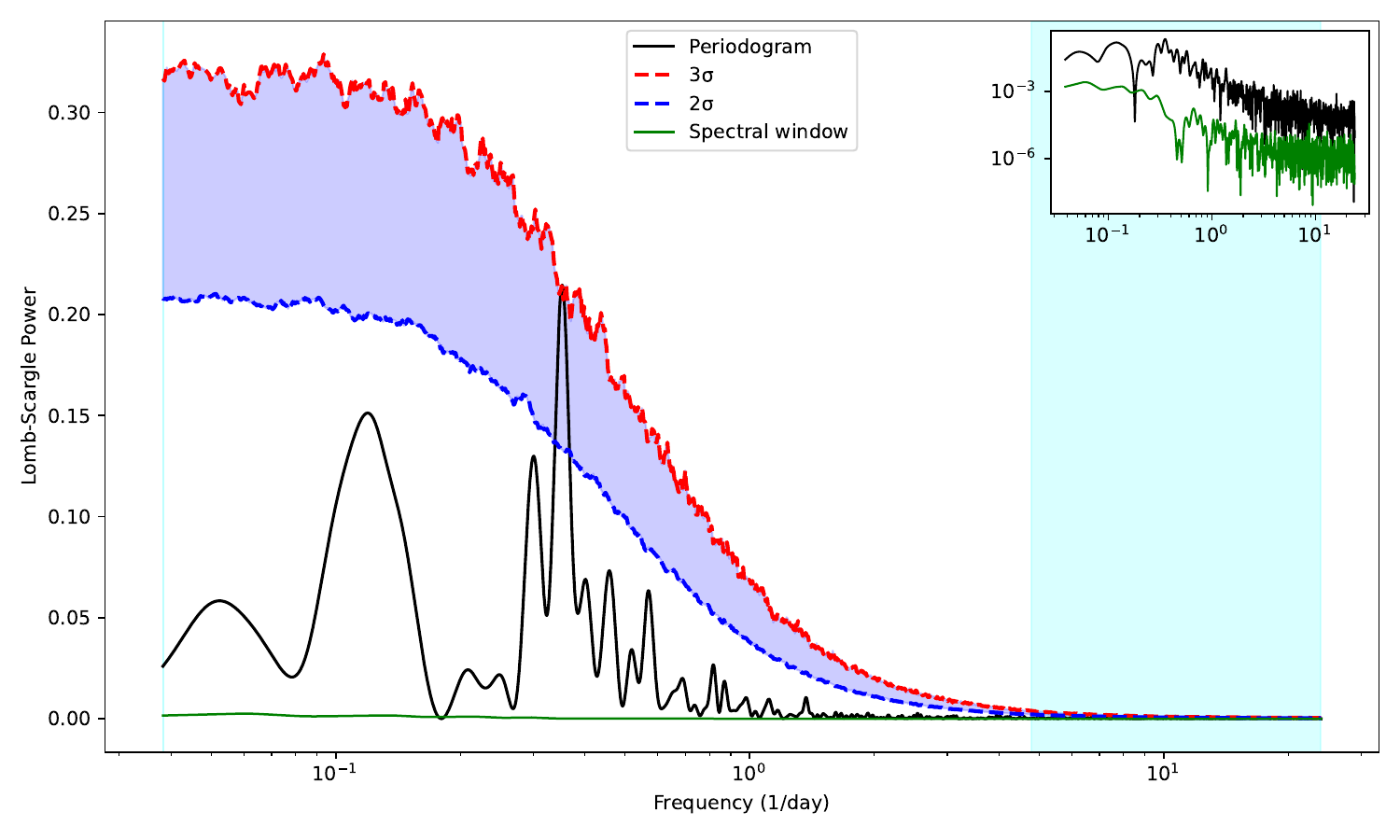}
    \caption{DRW-based significance assessment of the 
    generalized Lomb--Scargle periodogram for the 
    Sector~5 light curve of PKS~0521$-$36. The black 
    solid curve shows the observed LSP periodogram. 
    The red and blue dashed curves represent the 
    $3\sigma$ and $2\sigma$ confidence levels, 
    respectively, derived from an ensemble of 12\,000 
    mock light curves generated using the maximum-likelihood 
    DRW parameters inferred from the data via the 
    \textsc{EzTao} package. The 
    blue shaded region indicates the area between the 
    $2\sigma$ and $3\sigma$ thresholds. The green solid 
    curve shows the spectral window function, which 
    traces the power introduced by the uneven sampling 
    pattern of the TESS observations. The vertical cyan 
    line on the left marks the minimum frequency 
    corresponding to the total temporal baseline of the 
    sector ($f_{\rm min} = 1/T$), while the cyan shaded 
    region on the right delineates the frequencies 
    approaching the Nyquist limit ($f_{\rm max} = 
    1/2\Delta T$), beyond which the periodogram is 
    unreliable. The inset shows the full frequency range 
    on a logarithmic scale for both the periodogram 
    (black) and the spectral window (green). A prominent 
    peak near $f \approx 0.34$~d$^{-1}$ exceeds the 
    $3\sigma$ confidence threshold, providing independent 
    support for the QPO candidate identified in the LSP 
    analysis of Sector~5. The corresponding period and 
    significance are discussed in 
    Section~\ref{sec:results}.}
    \label{fig:drw_sig_s5}
\end{figure*}

\subsubsection{Damped Random Walk Significance Framework}
\label{sec:drw_sig}

As a complementary and independent significance test for the 
candidate QPO identified in Sector~5, we employed a damped 
random walk (DRW) based null-hypothesis framework. The DRW 
model corresponds to the simplest continuous autoregressive 
moving average process, CARMA(1,0), and has been widely 
adopted to characterize the stochastic red-noise variability 
of AGN light curves 
\citep{moreno2019stochastic, burke2021characteristic, 
zhang2022characterizing, zhang2023gaussian, 
sharma2024microquasars, TANTRY2025100372}. Its power spectral 
density takes the form of a broken power law, transitioning 
from a flat spectrum at low frequencies to a 
$P(\omega) \propto \omega^{-2}$ red-noise slope at high 
frequencies, making it a physically motivated null hypothesis 
against which periodic features can be tested.

The DRW parameters were inferred from the Sector~5 light 
curve using the publicly available \textsc{EzTao} 
package\footnote{\url{https://eztao.readthedocs.io/en/latest/}}, 
which is built on top of the \textsc{celerite} Gaussian-process 
framework\footnote{\url{https://celerite.readthedocs.io/en/stable/}} 
\citep{2013PASP..125..306F}. Parameter estimation was performed 
via Markov Chain Monte Carlo (MCMC) sampling implemented using 
the \texttt{emcee} 
package\footnote{\url{https://github.com/dfm/emcee}} within 
\textsc{EzTao}. These maximum-likelihood parameters were then 
used to generate 12\,000 mock light curves sharing the same 
stochastic properties as the observed data and sampled at 
identical time stamps. The generalized LSP was computed for 
each realization, and significance thresholds at the $1\sigma$, 
$2\sigma$, $3\sigma$, and $4\sigma$ confidence levels were 
derived at each trial frequency from the resulting ensemble. 
A spectral window periodogram was additionally constructed 
to identify aliasing features introduced by the uneven 
sampling, ensuring that any candidate QPO peak can be 
distinguished from sampling artifacts. The DRW-based 
significance curves and spectral window periodogram for 
Sector~5 are presented in Section~\ref{sec:results}.

\begin{table*}
    \centering
    \caption{Best-fitting PSD model parameters for the TESS 
    full hybrid light curves of PKS~0521$-$36 for Sectors~5, 
    6, and~32. For each sector, results are listed for the 
    simple power law ($M_1$), bending power law A ($M_2$), 
    and bending power law B ($M_3$). The preferred model 
    is selected on the basis of the minimum Bayesian 
    Information Criterion (BIC) value. Uncertainties 
    correspond to a change of $\Delta\mathcal{L} = 1$ 
    from the minimum of the log-likelihood statistic.}
    \label{tab:psd}
    \begin{tabular}{llccc}
    \hline\hline
    Model & Parameter & Sector~5 & Sector~6 & Sector~32 \\
    \hline
    \hline
    $M_1$ & $A$             
    & $1.302^{+0.053}_{-0.050}$ 
    & $0.548^{+0.025}_{-0.023}$ 
    & $1.724^{+0.130}_{-0.118}$ \\
    & $\alpha_1$             
    & $1.912^{+0.032}_{-0.032}$ 
    & $1.974^{+0.036}_{-0.036}$ 
    & $1.955^{+0.048}_{-0.047}$ \\
    & $c$                    
    & $0.0194^{+0.0007}_{-0.0007}$ 
    & $0.0053^{+0.0002}_{-0.0002}$ 
    & $0.0201^{+0.0004}_{-0.0004}$ \\
    & BIC                    
    & $-19384.1$ 
    & $-30793.6$ 
    & $-25078.8$ \\
    \hline
    $M_2$ & $A$             
    & $6.857^{+1.919}_{-1.264}$ 
    & $0.940^{+0.048}_{-0.077}$ 
    & $4.395^{+1.188}_{-0.799}$ \\
    & $\nu_b$ (d$^{-1}$)    
    & $0.308^{+0.103}_{-0.067}$ 
    & $1.568^{+0.157}_{-0.137}$ 
    & $0.851^{+0.258}_{-0.137}$ \\
    & $\alpha_1$             
    & $2.126^{+0.045}_{-0.058}$ 
    & $2.896^{+0.116}_{-0.109}$ 
    & $2.339^{+0.084}_{-0.101}$ \\
    & $c$                    
    & $0.0209^{+0.0007}_{-0.0007}$ 
    & $0.0070^{+0.0002}_{-0.0002}$ 
    & $0.0209^{+0.0004}_{-0.0004}$ \\
    & BIC                    
    & $-19399.9$ 
    & $-30940.2$ 
    & $-25098.7$ \\
    \hline
    $M_3$ & $A$             
    & $2.222^{+0.471}_{-0.197}$ 
    & $1.215^{+4.129}_{-0.052}$ 
    & $2.583^{+1.533}_{-0.168}$ \\
    & $\nu_b$ (d$^{-1}$)    
    & $1.856^{+0.672}_{-1.256}$ 
    & $1.194^{+0.022}_{-0.734}$ 
    & $1.993^{+0.238}_{-1.354}$ \\
    & $\alpha_1$             
    & $2.322$ 
    & $2.831^{+0.043}_{-0.602}$ 
    & $2.543$ \\
    & $\alpha_2$             
    & $1.549$ 
    & $0.588^{+0.052}_{-0.313}$ 
    & $1.290$ \\
    & $c$                    
    & $0.0210^{+0.0049}_{-0.0028}$ 
    & $0.0071^{+0.0005}_{-0.0013}$ 
    & $0.0210^{+0.0009}_{-0.0015}$ \\
    & BIC                    
    & $-19375.2$ 
    & $-30954.5$ 
    & $-25085.3$ \\
    \hline
    \multicolumn{5}{l}{\textit{Preferred model}} \\
    & Best fit               & $M_2$  & $M_3$  & $M_2$  \\
    \hline\hline
    \end{tabular}
    \begin{minipage}{\textwidth}
    \vspace{0.2cm}
    \small
    Note: For $M_3$ parameters $\alpha_1$ and $\alpha_2$ 
    in Sectors~5 and~32, profile likelihood uncertainties 
    could not be reliably determined due to the broad 
    likelihood surface near the minimum, and only the 
    best-fitting values are reported.
    \end{minipage}
\end{table*}

\begin{figure*}
    \centering
    \begin{subfigure}{0.65\textwidth}
        \centering
        \includegraphics[width=\textwidth]{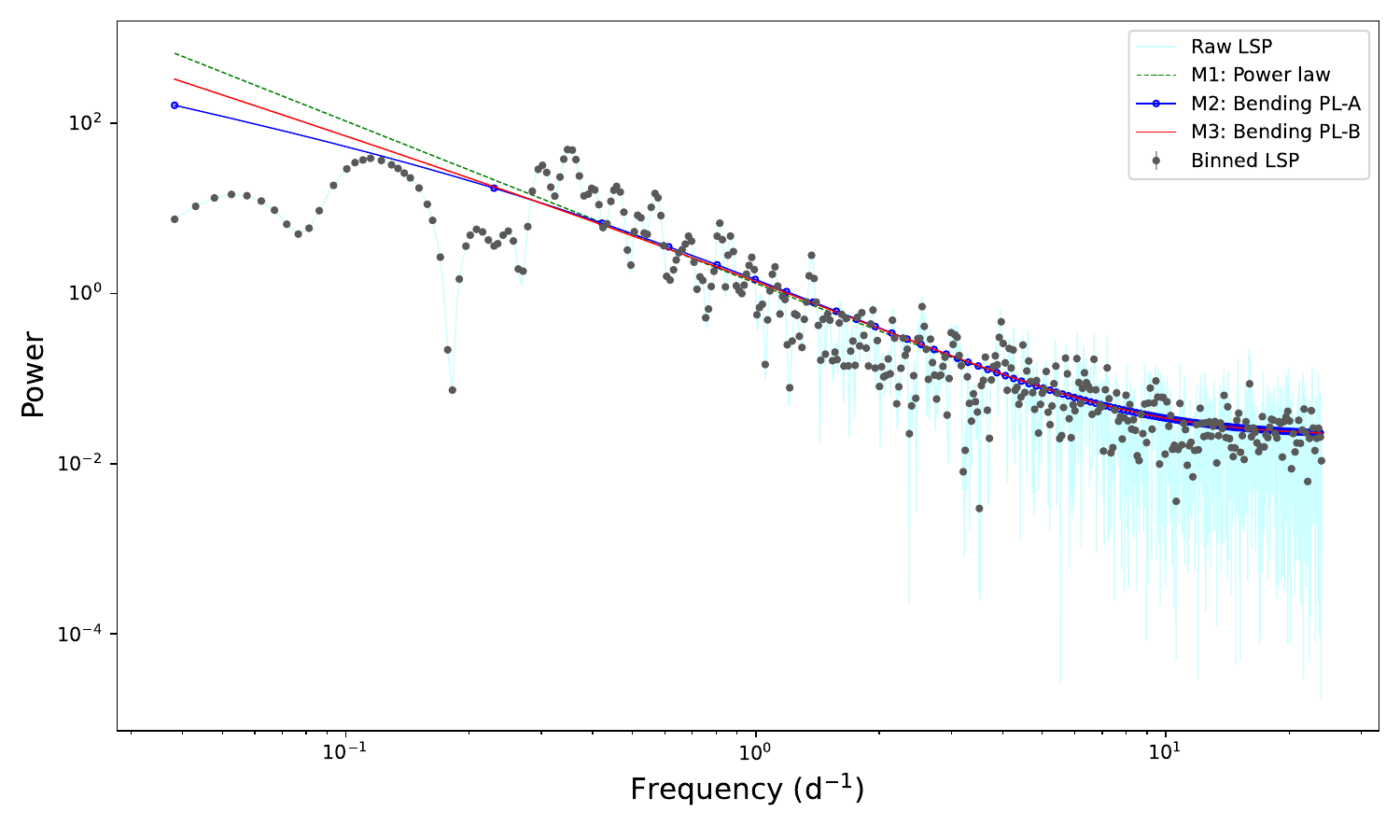}
        \caption{Sector~5: preferred model $M_2$ (bending 
        power law-A) with bending frequency 
        $\nu_b = 0.308^{+0.103}_{-0.067}$~d$^{-1}$.}
        \label{fig:psd_s5}
    \end{subfigure}

    \vspace{0.1cm}

    \begin{subfigure}{0.65\textwidth}
        \centering
        \includegraphics[width=\textwidth]{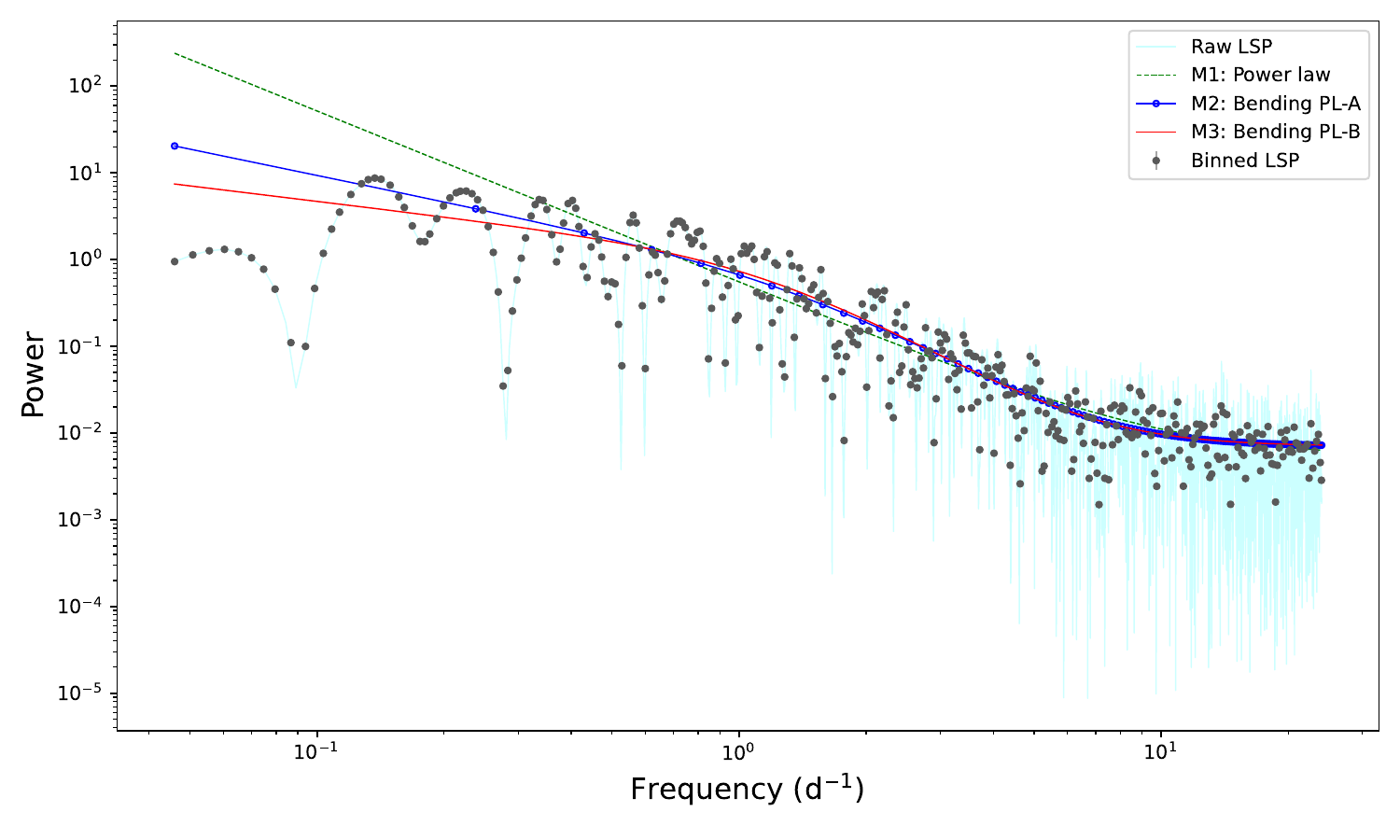}
        \caption{Sector~6: preferred model $M_3$ (bending 
        power law-B) with bending frequency 
        $\nu_b = 1.194^{+0.023}_{-0.734}$~d$^{-1}$.}
        \label{fig:psd_s6}
    \end{subfigure}

    \vspace{0.1cm}

    \begin{subfigure}{0.65\textwidth}
        \centering
        \includegraphics[width=\textwidth]{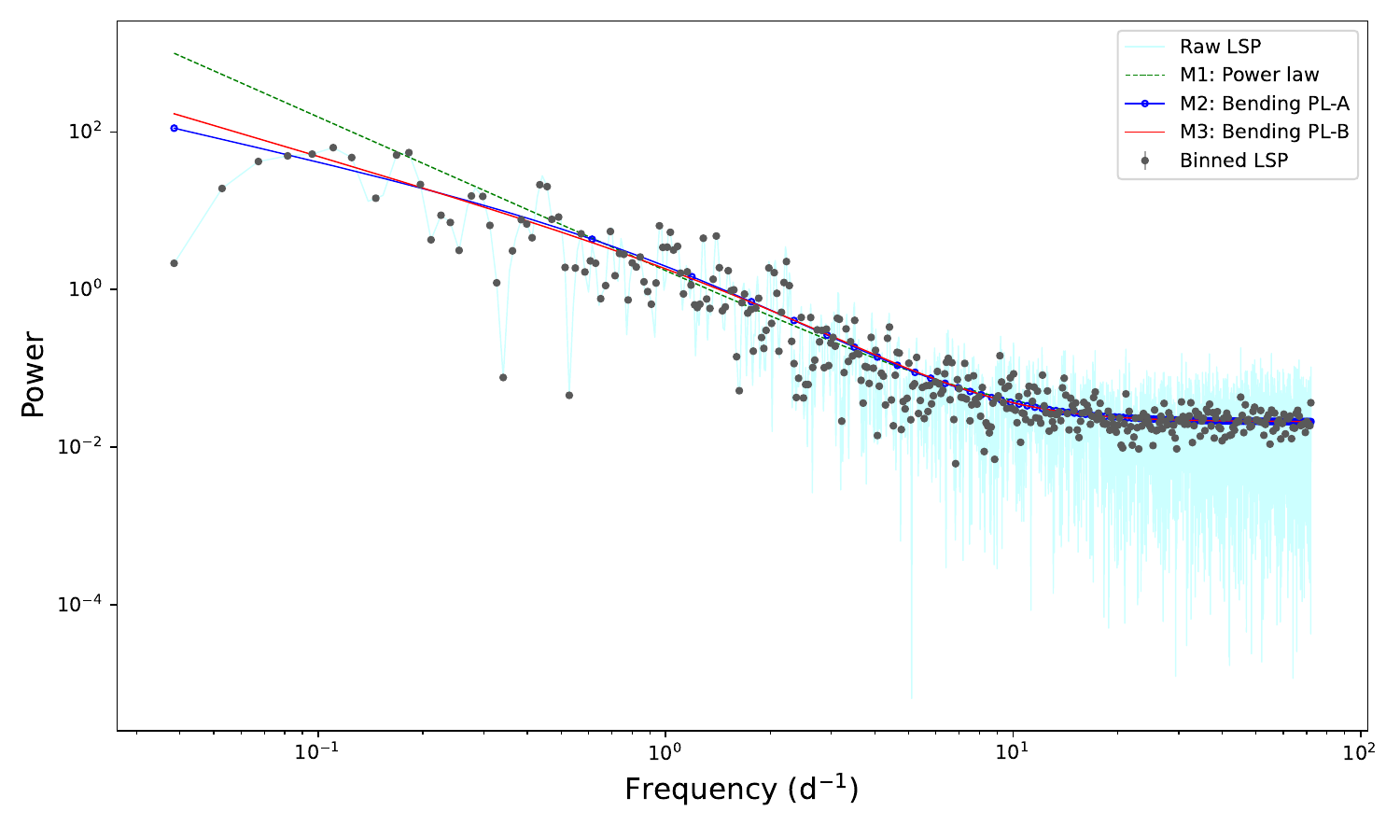}
        \caption{Sector~32: preferred model $M_2$ (bending 
        power law-A) with bending frequency 
        $\nu_b = 0.851^{+0.258}_{-0.137}$~d$^{-1}$.}
        \label{fig:psd_s32}
    \end{subfigure}

    \caption{Power spectral density analysis of the TESS 
    full hybrid light curves of PKS~0521$-$36 for 
    Sectors~5, 6, and~32. In each panel, the cyan points 
    with error bars show the logarithmically rebinned 
    periodogram, and the light cyan curve shows the raw 
    Lomb--Scargle periodogram. The green dashed, blue 
    solid, and red solid curves correspond to the 
    best-fitting $M_1$ (simple power law), $M_2$ (bending 
    power law-A), and $M_3$ (bending power law-B) models, 
    respectively. The preferred model in each sector, 
    selected on the basis of the minimum BIC value, is 
    listed in the subcaption. Best-fitting parameters for 
    all models are given in Table~\ref{tab:psd}.}
    \label{fig:psd_all}
\end{figure*}

\subsection{Periodogram Analysis}
\label{sec:psd}

The optical light curves of AGNs typically exhibit 
red-noise variability whose power spectral density (PSD) follows 
a power-law form $P(\nu) \sim \nu^{-\alpha}$, where $\nu$ is the 
temporal frequency and the power decreases monotonically with 
increasing frequency. In addition to this simple power-law 
behavior, bending or broken power laws have also been observed 
to characterize AGN PSD shapes \citep{2012A&A...544A..80G}, 
with the break frequency providing a physically meaningful 
characteristic timescale of the system. Fitting the PSD with 
appropriate models therefore allows both the spectral slope 
and any characteristic variability timescale to be constrained, 
and any oscillatory feature present in the light curve can be 
tested for significance against the underlying continuum model.

In this work, we computed the generalized Lomb--Scargle 
periodogram \citep[LSP;][]{1976Ap&SS..39..447L, 
1982ApJ...263..835S, 
2009A&A...496..577Z, 2018ApJS..236...16V} for each sector 
individually using the \textsc{LombScarglePowerspectrum} 
implementation from the \textsc{Stingray} 
package\footnote{\url{https://docs.stingray.science/}}, 
incorporating the measured flux uncertainties. The frequency 
grid was constructed over the range $f_{\rm min} = 1/T$ to 
$f_{\rm max} = 1/(2\Delta T)$, where $T$ is the total temporal 
baseline of each sector and $\Delta T$ is the median sampling 
interval. The resulting periodogram was subsequently rebinned 
on a logarithmic frequency grid to reduce scatter and improve 
the visual clarity of the spectral shape.

The PSD was fitted with three distinct spectral models 
\citep[e.g.,][]{2010MNRAS.402..307V, 2013MNRAS.433..907E}:

\noindent $M_1$ (simple power law):
\begin{equation}
P(\nu) = A\nu^{-\alpha_1} + c,
\end{equation}

\noindent $M_2$ (bending power law -- `A'):
\begin{equation}
P(\nu) = A\nu^{-1}
\left[1 + \left(\frac{\nu}{\nu_b}\right)^{\alpha_1 - 1}
\right]^{-1} + c,
\end{equation}

\noindent and $M_3$ (bending power law -- `B'):
\begin{equation}
P(\nu) = A\nu^{-\alpha_2}
\left[1 + \left(\frac{\nu}{\nu_b}\right)^{\alpha_1 - \alpha_2}
\right]^{-1} + c,
\end{equation}

\noindent where $A$, $\alpha_1$, $\alpha_2$, $\nu_b$, and $c$ 
are the normalization, spectral indices, bending frequency, and 
an additive constant, respectively. Models $M_1$, $M_2$, and 
$M_3$ have three, four, and five free parameters, respectively.

The best-fitting parameters for each model were obtained by 
minimizing the negative log-likelihood statistic 
\citep{2010MNRAS.402..307V}
\begin{equation}
\mathcal{L} = -2\sum_{j} 
\frac{I_j}{P_j} + \log P_j,
\end{equation}
where $I_j$ and $P_j$ are the observed periodogram power and 
the model spectrum at frequency $\nu_j$, respectively. 
Minimization was performed using the Limited-memory 
Broyden--Fletcher--Goldfarb--Shanno with Bound constraints 
(L-BFGS-B) algorithm, implemented via the 
\texttt{scipy.optimize.minimize} routine from the 
\textsc{SciPy} package \citep{2020NatMe..17..261V}, and 
parameter uncertainties were estimated through profile 
likelihood analysis following \citet{2012A&A...544A..80G}. 
In this approach, each parameter is varied individually 
while the remaining parameters are re-optimized at each 
step, and the $1\sigma$ confidence interval is defined 
by the change $\Delta\mathcal{L} = 1$ from the minimum 
of the likelihood statistic.

To select the preferred model among $M_1$, $M_2$, and $M_3$, 
we computed the Bayesian Information Criterion (BIC) for each 
fit, defined as
\begin{equation}
\mathrm{BIC} = k \cdot \ln(n) - \mathcal{L},
\end{equation}
where $k$ is the number of free parameters and $n$ is the 
number of frequency bins. The model with the lowest BIC value 
is taken as the preferred description of the PSD shape. The 
best-fitting parameters and BIC values for all three sectors 
are presented in Table~\ref{tab:psd}, and the results are 
discussed in Section~\ref{sec:results}.

\begin{figure*}
    \centering
    \begin{subfigure}{\textwidth}
        \centering
        \includegraphics[width=0.45\textwidth]{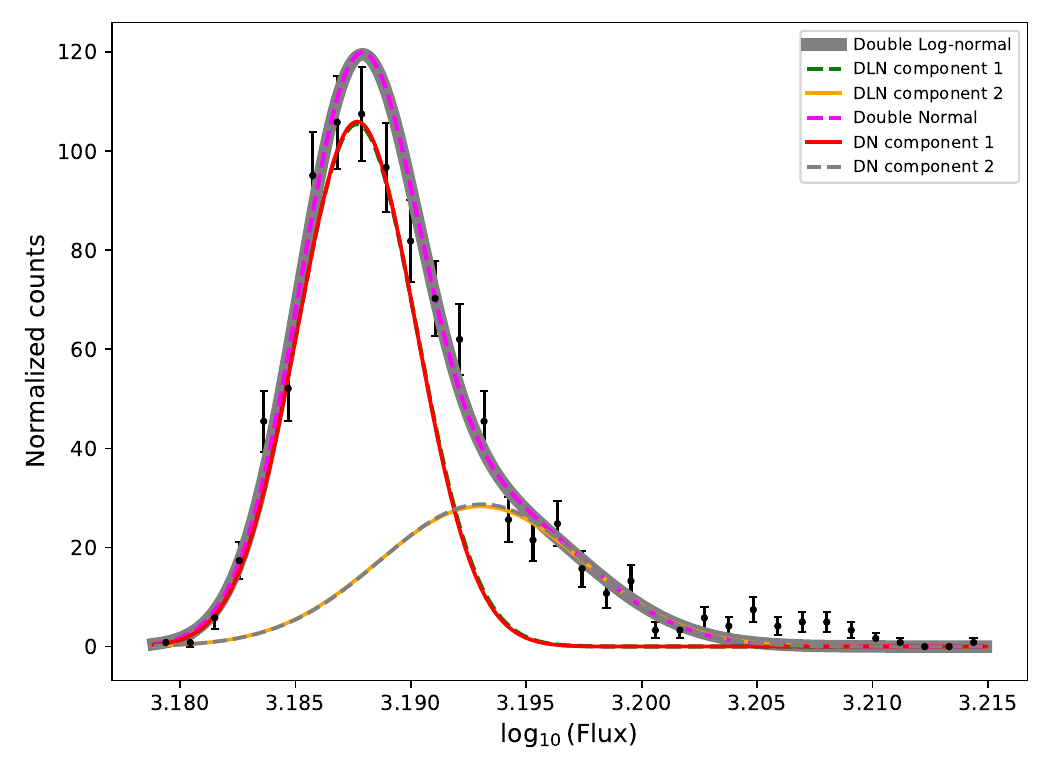}
        \caption{Sector~5 (1437.99--1464.14 BTJD). The 
        distribution peaks near 
        $\log_{10}(\mathrm{Flux}) \approx 3.188$ and displays 
        a pronounced asymmetric tail toward higher flux values, 
        indicative of sporadic flaring activity consistent with 
        the light curve.}
        \label{fig:fd_s5}
    \end{subfigure}
    
    \vspace{0.1cm}
    
    \begin{subfigure}{\textwidth}
        \centering
        \includegraphics[width=0.45\textwidth]{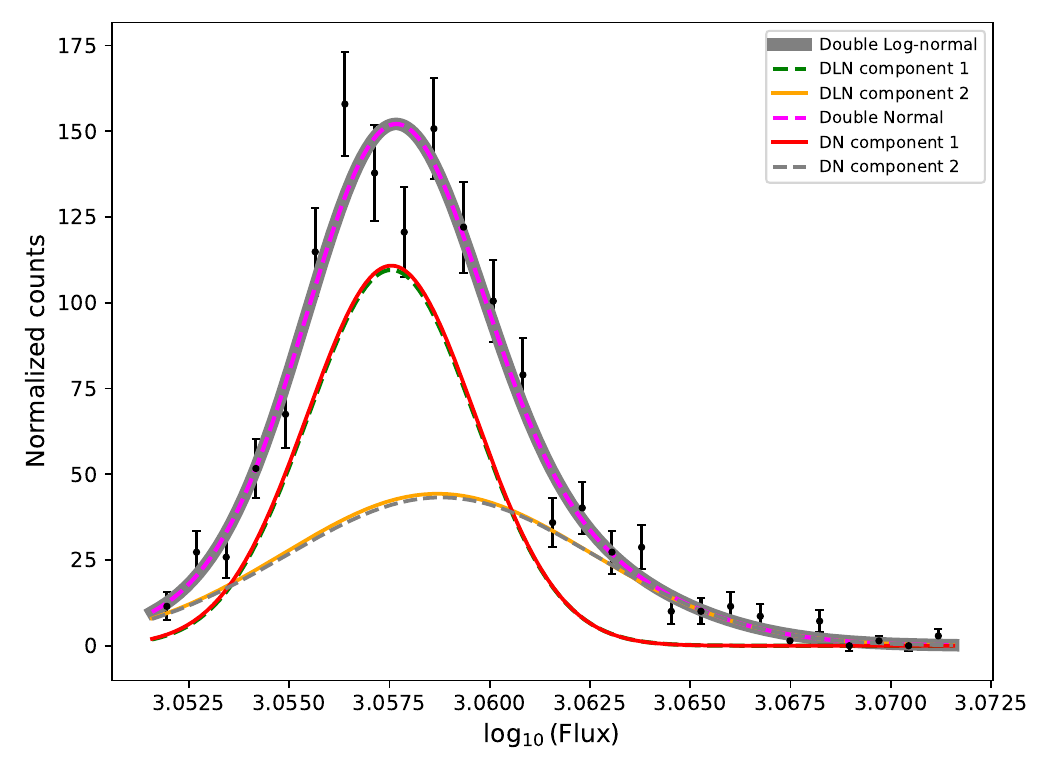}
        \caption{Sector~6 (1468.99--1490.03 BTJD). The 
        distribution peaks near 
        $\log_{10}(\mathrm{Flux}) \approx 3.058$ and exhibits 
        a clearly asymmetric profile with an extended high-flux 
        tail, requiring a two-component model for an adequate 
        description.}
        \label{fig:fd_s6}
    \end{subfigure}
    
    \vspace{0.1cm}
    
    \begin{subfigure}{\textwidth}
        \centering
        \includegraphics[width=0.45\textwidth]{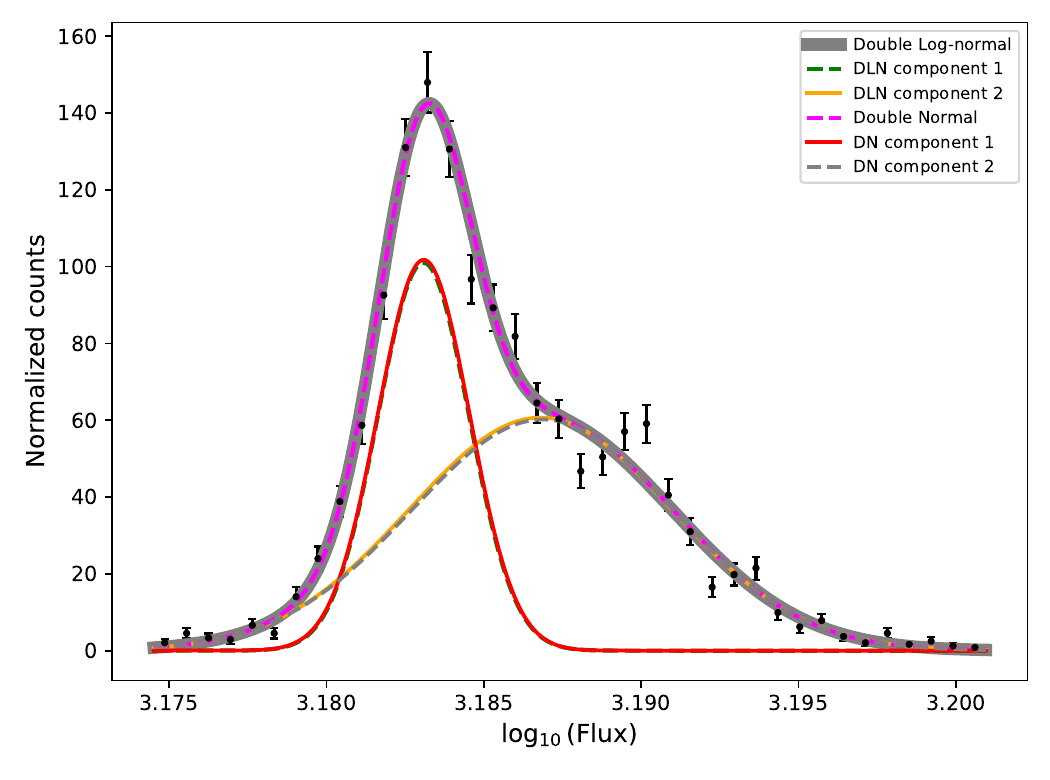}
        \caption{Sector~32 (2174.24--2200.23 BTJD). The 
        distribution peaks near 
        $\log_{10}(\mathrm{Flux}) \approx 3.183$ and shows a 
        relatively symmetric core with a moderate high-flux 
        tail. The double log-normal and double normal fits are 
        nearly indistinguishable in this sector.}
        \label{fig:fd_s32}
    \end{subfigure}
    
    \caption{Normalized flux distributions of the TESS full 
hybrid light curves of PKS~0521$-$36 for Sectors~5, 6, 
and~32, plotted as a function of $\log_{10}(\mathrm{Flux})$. 
Black points with error bars represent the observed 
normalized histogram counts. The thick gray solid curve 
shows the best-fitting double log-normal (DLN) model, with 
its two individual components shown as green dashed and 
orange solid curves, respectively. The magenta dashed curve 
shows the best-fitting double normal (DN) model, with its 
two components shown as red solid and gray dashed curves, 
respectively. In all three sectors, neither a single normal 
nor a single log-normal function provides an adequate 
representation of the observed distribution, as indicated 
by the Anderson--Darling test, and two-component models are 
required. The best-fitting parameters and the corresponding 
reduced $\chi^2$ values are listed in 
Table~\ref{tab:flux_dist}.}
    \label{fig:flux_dist_all}
\end{figure*}

\begin{table*}
    \centering
    \caption{Flux distribution statistics for the TESS full 
    hybrid light curves of PKS~0521$-$36 for Sectors~5, 6, 
    and~32. The skewness of the flux distribution, the 
    Anderson--Darling (AD) statistics for both the Gaussian 
    and log-normal single-component fits, and the 
    best-fitting parameters with reduced $\chi^2_\nu$ values 
    for the double log-normal (DLN) and double normal (DN) 
    two-component models are listed. The 5\% critical value 
    for the AD test is 0.786, which is exceeded by all 
    sectors, firmly rejecting single-component models.}
    \label{tab:flux_dist}
    \begin{tabular}{lccc}
    \hline\hline
    Quantity & Sector~5 & Sector~6 & Sector~32 \\
    \hline
    \multicolumn{4}{l}{\textit{Basic statistics}} \\
    Skewness (linear flux)       & 1.50   & 0.78   & 0.68   \\
    Skewness (log flux)          & 1.45   & 0.76   & 0.66   \\
    \hline
    \multicolumn{4}{l}{\textit{Anderson--Darling test}} \\
    Gaussian AD statistic        & 30.76  & 6.00   & 45.34  \\
    Log-normal AD statistic      & 29.21  & 5.66   & 43.56  \\
    \hline
    \multicolumn{4}{l}{\textit{Double log-normal (DLN) fit}} \\
    $a$                 
    & $0.685 \pm 0.163$ 
    & $0.573 \pm 0.163$ 
    & $0.366 \pm 0.021$ \\
    $\sigma_1$          
    & $0.00259 \pm 0.00025$ 
    & $0.00209 \pm 0.00027$ 
    & $0.00145 \pm 0.00007$ \\
    $\mu_1$             
    & $3.1877 \pm 0.0002$ 
    & $3.0575 \pm 0.0002$ 
    & $3.1831 \pm 0.0001$ \\
    $\sigma_2$          
    & $0.00444 \pm 0.00095$ 
    & $0.00384 \pm 0.00038$ 
    & $0.00417 \pm 0.00008$ \\
    $\mu_2$             
    & $3.1930 \pm 0.0024$ 
    & $3.0587 \pm 0.0006$ 
    & $3.1868 \pm 0.0001$ \\
    $\chi^2_\nu$        & 2.31   & 1.52   & 2.40   \\
    \hline
    \multicolumn{4}{l}{\textit{Double normal (DN) fit}} \\
    $a$                 
    & $0.684 \pm 0.160$ 
    & $0.582 \pm 0.160$ 
    & $0.371 \pm 0.021$ \\
    $\sigma_1$\,(e$^-$\,s$^{-1}$)          
    & $9.14 \pm 0.88$   
    & $5.50 \pm 0.69$   
    & $5.11 \pm 0.25$   \\
    $\mu_1$\,(e$^-$\,s$^{-1}$)  
    & $1540.5 \pm 0.7$  
    & $1141.7 \pm 0.5$  
    & $1524.3 \pm 0.2$  \\
    $\sigma_2$\,(e$^-$\,s$^{-1}$)          
    & $15.77 \pm 3.35$  
    & $10.18 \pm 0.99$  
    & $14.76 \pm 0.27$  \\
    $\mu_2$\,(e$^-$\,s$^{-1}$)  
    & $1559.7 \pm 8.4$  
    & $1144.9 \pm 1.8$  
    & $1537.6 \pm 0.5$  \\
    $\chi^2_\nu$        & 2.36   & 1.52   & 2.41   \\
    \hline
    \multicolumn{4}{l}{\textit{Preferred model}} \\
    Best fit            & DLN    & DLN$^a$    & DLN    \\
    \hline\hline
    \end{tabular}
    \begin{minipage}{\textwidth}
    \vspace{0.2cm}
    \small
     $^a$ In Sector~6, the double log-normal and double normal 
models yield nearly identical reduced $\chi^2_\nu$ values 
of 1.5162 and 1.5161, respectively, with the double normal 
being marginally preferred on the basis of $\chi^2_\nu$ 
alone. Nevertheless, the double log-normal model is adopted 
as the preferred description for consistency with the other 
sectors, and given that the difference of $\Delta\chi^2_\nu 
= 0.0001$ is negligible and carries no physical 
significance.
    \end{minipage}
\end{table*}

\subsection{Flux Distribution}
\label{sec:flux_dist}

The statistical properties of the flux distribution provide 
important insights into the nature of the underlying variability 
process in AGN emission. A log-normal flux distribution is 
commonly observed in blazars and AGNs across multiple wavelength 
bands \citep{uttley2005non, shah2018log, 
10.1093/mnrasl/sly136, 10.1093/mnras/stz3108,2026MNRAS.545f1920T,2021ApJ...919...58Z}, and is generally associated 
with multiplicative processes in the accretion disc or jet 
\citep{Uttley_2001,2025MNRAS.539.2185M, 2024ApJ...977..111A}. If the disc and jet variability are 
coupled, fluctuations originating in the disc can propagate 
into the jet and imprint a log-normal signature on the observed 
flux distribution \citep{2024MNRAS.527.2672S}.

To investigate the statistical behavior of the optical 
variability in each TESS sector, we examined the flux 
distribution of the PKS~0521$-$36 light curves using the 
Anderson--Darling (AD) normality test together with histogram 
fitting. The AD test was applied to both the linear flux values 
and their logarithms to assess whether the distributions are 
consistent with a single normal or single log-normal model. 
When the AD statistic exceeds the 5\% critical value of 0.786, 
the null hypothesis of normality is rejected, indicating that 
a more complex description of the PDF is required.

In cases where neither a single normal nor a single log-normal 
model provides an adequate representation of the observed 
distribution, we constructed a normalized histogram of the 
logarithm of the flux and fitted it using two-component models: 
a double log-normal and a double normal function. The double 
log-normal model is defined as
\begin{equation}
\label{eq:double_lognorm}
\begin{split}
D_{\mathrm{dLN}}(x) 
&= a\,\frac{1}{\sqrt{2\pi}\,\sigma_{1}}
   \exp\!\left[-\frac{(x-\mu_{1})^{2}}{2\sigma_{1}^{2}}\right] \\
&\quad + (1-a)\,\frac{1}{\sqrt{2\pi}\,\sigma_{2}}
   \exp\!\left[-\frac{(x-\mu_{2})^{2}}{2\sigma_{2}^{2}}\right],
\end{split}
\end{equation}
where $a$ is the mixing fraction, $\mu_1$ and $\mu_2$ are the 
centroids of the two components, and $\sigma_1$ and $\sigma_2$ 
are their corresponding widths. The double normal model, 
defined in the linear flux domain and evaluated in logarithmic 
space, takes the form
\begin{equation}
\label{eq:double_norm}
D_{\rm dN}(x) =
a\,\mathcal{N}(x;\sigma_{1},\mu_{1}) + 
(1-a)\,\mathcal{N}(x;\sigma_{2},\mu_{2}),
\end{equation}
with
\begin{equation}
\mathcal{N}(x;\sigma,\mu) =
\frac{1}{\sqrt{2\pi}\,\sigma}
\exp\!\left[-\frac{(10^{x}-\mu)^{2}}{2\sigma^{2}}\right]
\,10^{x}\ln 10.
\end{equation}
The preferred model between the double log-normal and double 
normal descriptions was selected on the basis of the reduced 
$\chi^2$ statistic, with the model yielding the lower value 
taken as the better representation of the observed flux 
distribution. The best-fitting parameters and the 
corresponding reduced $\chi^2$ values for each sector are 
reported in Table~\ref{tab:flux_dist}, and the flux 
distribution histograms with the fitted models are shown 
in Section~\ref{sec:results}.

\section{Results}
\label{sec:results}

In this section, we present the results of the time-series 
analyses described in Section~\ref{sec:analysis}, organized 
sector-wise for the three TESS observations of PKS~0521$-$36. 
For each sector, we report the fractional variability 
amplitude, the PSD model fit, the flux distribution 
properties, and the outcome of the QPO search using the 
LSP and WWZ methods.

\subsection{Sector 5}
\label{sec:res_s5}

\subsubsection{Fractional Variability}

The Sector~5 light curve (1437.99--1464.14 BTJD; 
15~November--11~December~2018) spans approximately 26.1~days 
at 30~min cadence. The source 
exhibits a flux range of 1509.6--1640.2~e$^-$\,s$^{-1}$, 
with a positive skewness of 1.50, the highest among the three 
sectors, consistent with the prominent flaring activity visible 
in the light curve (Figure~\ref{fig:tess_lc}). The fractional 
variability amplitude is $F_{\rm var} = (1.19 \pm 0.002)\%$, 
the largest value measured across all three sectors, further 
confirming the elevated level of optical activity during this 
epoch.

\subsubsection{Power Spectral Density}

The PSD of Sector~5 is best described by the bending 
power-law model $M_2$, which yields the lowest BIC value 
of $-19399.9$ among the three models considered 
(Table~\ref{tab:psd}). The best-fitting parameters are 
a normalization $A = 6.857^{+1.919}_{-1.264}$, a bending 
frequency $\nu_b = 0.308^{+0.103}_{-0.067}$~d$^{-1}$ 
corresponding to a characteristic timescale of 
$\sim$3.2~d, and a high-frequency spectral index 
$\alpha_1 = 2.126^{+0.045}_{-0.058}$. The preference for 
a bending power law over a simple power law indicates 
the presence of a characteristic variability timescale 
within the temporal baseline of this sector.

\subsubsection{Flux Distribution}

The flux distribution of Sector~5 is strongly asymmetric, 
with a pronounced tail toward higher flux values 
(Figure~\ref{fig:flux_dist_all}). Both the Gaussian and 
log-normal Anderson--Darling statistics (30.76 and 29.21, 
respectively) far exceed the 5\% critical value of 0.786, 
firmly rejecting single-component models. The double 
log-normal model provides the best description of the 
observed distribution with a reduced $\chi^2_\nu = 2.31$, 
marginally better than the double normal fit 
($\chi^2_\nu = 2.36$). The two DLN components are centred 
at $\mu_1 = 3.1877$ and $\mu_2 = 3.1930$ in 
$\log_{10}(\mathrm{Flux})$, with a mixing fraction 
$a = 0.685$, indicating that the dominant component 
accounts for approximately 68\% of the flux states. The 
presence of two distinct flux components is consistent 
with the source alternating between a quiescent state 
and an elevated activity state during this sector.

\subsubsection{QPO Search}

The LSP of Sector~5 reveals a dominant peak at a frequency 
of $f = 0.3524 \pm 0.0097$~d$^{-1}$, corresponding to a 
period of $P = 2.838 \pm 0.078$~d (Figure~\ref{fig:lsp_all}). 
This peak exceeds the 99.99\% confidence level derived from 
the Monte Carlo simulations, as indicated by the filled 
purple circle in the figure. The DRW-based significance 
framework provides independent confirmation of this feature: 
the LSP peak at the same frequency exceeds the $3\sigma$ 
confidence envelope constructed from 12\,000 DRW-simulated 
light curves (Figure~\ref{fig:drw_sig_s5}), confirming that 
the signal cannot be attributed to red-noise fluctuations 
alone.

To assess the temporal persistence of the candidate QPO, 
we performed a WWZ analysis on the Sector~5 light curve 
binned to 4~hr resolution. The time--frequency map reveals 
a localized concentration of power near the same frequency, 
and the time-averaged WWZ spectrum yields a peak at 
$f = 0.3522 \pm 0.0137$~d$^{-1}$, corresponding to a 
period of $P = 2.839 \pm 0.110$~d, in excellent agreement 
with the LSP result. The consistency between the LSP and 
WWZ detections, combined with the  statistical 
significance established through both Monte Carlo and DRW 
frameworks, supports the interpretation of a 
quasi-periodic oscillation with a period of 
$\sim$2.84~d in the optical light curve of PKS~0521$-$36 
during Sector~5. The QPO is observed over approximately 
9 cycles within the 26.1-day baseline of the sector.

\subsection{Sector 6}
\label{sec:res_s6}

\subsubsection{Fractional Variability}

The Sector~6 light curve (1468.99--1490.03 BTJD; 
12~December~2018--6~January~2019) spans approximately 
21.1~days at 30~min cadence. 
The source flux ranges from 1126.1 to 1179.1~e$^-$\,s$^{-1}$, 
notably lower than in Sectors~5 and~32, suggesting that the 
source was in a comparatively fainter optical state during 
this epoch. The fractional variability amplitude is 
$F_{\rm var} = (0.69 \pm 0.003)\%$, the lowest among 
the three sectors.

\subsubsection{Power Spectral Density}

The PSD of Sector~6 is best described by the bending 
power-law model $M_3$, which yields the lowest BIC value 
of $-30954.5$ (Table~\ref{tab:psd}). The best-fitting 
parameters are a bending frequency 
$\nu_b = 1.194^{+0.023}_{-0.734}$~d$^{-1}$, a 
high-frequency spectral index 
$\alpha_1 = 2.831^{+0.043}_{-0.602}$, and a low-frequency 
index $\alpha_2 = 0.588^{+0.052}_{-0.313}$. The preference 
for the more complex $M_3$ model suggests that the PSD 
shape in this sector exhibits a more pronounced transition 
between the low- and high-frequency regimes compared to 
the other sectors.

\subsubsection{Flux Distribution}

The flux distribution of Sector~6 shows a moderately 
asymmetric profile with a positive skewness of 0.78 
and an extended high-flux tail (Figure~\ref{fig:flux_dist_all}). 
The AD statistics for the Gaussian (6.00) and log-normal 
(5.66) fits both exceed the 5\% critical value, rejecting 
single-component models. The double log-normal and double 
normal models yield nearly identical reduced $\chi^2_\nu$ 
values of 1.52 and 1.52, respectively, making them 
statistically indistinguishable. The DLN model is adopted 
as the preferred description for consistency with the 
other sectors, with the two components centred at 
$\mu_1 = 3.0575$ and $\mu_2 = 3.0587$ and a mixing 
fraction $a = 0.573$.

\subsubsection{QPO Search}

The LSP of Sector~6 does not reveal any peak exceeding 
the 97.5\%, 99.7\%, or 99.9\% Monte Carlo confidence 
levels across the explored frequency range 
(Figure~\ref{fig:lsp_all}). No statistically significant 
periodic feature is therefore identified in this sector, 
and the WWZ and DRW analyses are not pursued further for 
Sector~6.

\subsection{Sector 32}
\label{sec:res_s32}

\subsubsection{Fractional Variability}

The Sector~32 light curve (2174.24--2200.23 BTJD; 
19~November--16~December~2020) spans approximately 
26.0~days at 10~min cadence, 
providing the densest temporal sampling of the three 
sectors. The source flux ranges from 1494.5 to 
1588.4~e$^-$\,s$^{-1}$, comparable to Sector~5, and the 
fractional variability amplitude is 
$F_{\rm var} = (0.80 \pm 0.003)\%$, intermediate between 
the values measured in Sectors~5 and~6.

\subsubsection{Power Spectral Density}

The PSD of Sector~32 is best described by the bending 
power-law model $M_2$, with a minimum BIC value of 
$-25098.7$ (Table~\ref{tab:psd}). The best-fitting 
bending frequency is 
$\nu_b = 0.851^{+0.258}_{-0.137}$~d$^{-1}$, corresponding 
to a characteristic timescale of $\sim$1.2~d, and the 
high-frequency spectral index is 
$\alpha_1 = 2.339^{+0.084}_{-0.101}$. The bending 
frequency in Sector~32 is higher than that found in 
Sector~5, suggesting that the dominant variability 
timescale shifted to shorter periods between the two 
Cycle~1 and Cycle~3 epochs.

\subsubsection{Flux Distribution}

The flux distribution of Sector~32 shows a relatively 
symmetric core with a moderate high-flux tail and a 
positive skewness of 0.68 
(Figure~\ref{fig:flux_dist_all}). The AD statistics 
for both the Gaussian (45.34) and log-normal (43.56) 
fits are the largest among the three sectors, strongly 
rejecting single-component models despite the relatively 
modest visual asymmetry. The double log-normal model 
provides the best fit with $\chi^2_\nu = 2.40$, 
marginally better than the double normal 
($\chi^2_\nu = 2.41$). The two DLN components are 
centred at $\mu_1 = 3.1831$ and $\mu_2 = 3.1868$ 
with a mixing fraction $a = 0.366$, indicating that 
the secondary component is more prominent in this 
sector relative to Sectors~5 and~6.

\subsubsection{QPO Search}

The LSP of Sector~32 does not reveal any peak exceeding 
the 97.5\% or 99.9\% Monte Carlo confidence levels 
(Figure~\ref{fig:lsp_all}). No statistically significant 
periodic feature is identified in this sector, and 
accordingly the WWZ and DRW significance analyses are 
not applied to Sector~32.

\subsection{Summary of Results}
\label{sec:res_summary}

The results of the sector-wise analysis are summarized 
in Tables~\ref{tab:fvar}, \ref{tab:psd}, and 
\ref{tab:flux_dist}. Across all three sectors, the 
fractional variability amplitude ranges from 0.69\% 
to 1.19\%, with Sector~5 consistently showing the 
highest level of optical activity. The PSD shapes are 
well described by bending power-law models in all 
sectors, with high-frequency spectral indices in the 
range $\alpha_1 \approx 2.1$--$2.9$, characteristic 
of red-noise dominated variability. The flux 
distributions in all sectors require two-component 
models, with the double log-normal providing the best 
or equally good description in each case.

A statistically significant QPO with a period of 
$P \approx 2.84$~d is detected in Sector~5 at 
$>99.99\%$ confidence in the LSP analysis and at 
$>3\sigma$ significance in the DRW-based framework, 
with the WWZ analysis yielding a consistent period 
of $2.839 \pm 0.110$~d from the 4~hr binned light 
curve. No significant periodic feature is found in 
Sectors~6 or~32. The physical implications of the 
detected QPO are discussed in 
Section~\ref{sec:sum}.

\section{Summary and Discussion}
\label{sec:sum}

We have presented a comprehensive analysis of high-cadence optical 
light curves of the non-blazar AGN PKS~0521$-$36 obtained by the 
Transiting Exoplanet Survey Satellite across three sectors: Sector~5 
(15 November--11 December 2018, 30~min cadence), Sector~6 
(12 December 2018--6 January 2019, 30~min cadence), and Sector~32 
(19 November--16 December 2020, 10~min cadence). Using the 
\textsc{Quaver} pipeline for systematics correction, we characterize 
the short-timescale optical variability through fractional variability 
estimation, power spectral density analysis, quasi-periodic oscillation 
searches employing the Lomb--Scargle periodogram and the weighted 
wavelet Z-transform, and statistical flux distribution analysis. The 
main results are summarized as follows.

\subsection*{Fractional Variability}

The source exhibits moderate optical variability across all three 
sectors, with $F_\mathrm{var}$ ranging from $0.69\%$ in Sector~6 to 
$1.19\%$ in Sector~5, with Sector~32 intermediate at $0.80\%$. The 
highest variability amplitude in Sector~5 is consistent with the 
prominent flaring activity visible in that sector's light curve and 
its positive skewness of 1.50.
 The relatively low $F_\mathrm{var}$ 
values compared to typical blazars are consistent with the moderately 
beamed nature of PKS~0521$-$36, where Doppler enhancement is less 
extreme than in classical blazars \citep{2019A&A...627A.148A}. The 
sector-to-sector variation in $F_\mathrm{var}$ indicates that the 
optical activity level of the source changes on timescales of months 
to years, with Sector~6 representing a comparatively quiescent epoch.

\subsection*{Power Spectral Density}

The PSD shapes across all three sectors are better described by 
bending power-law models than by a simple power law. Sectors~5 and~32 
are best fit by model $M_2$ (bending power law A), with bending 
frequencies of $\nu_b = 0.308^{+0.103}_{-0.067}$~d$^{-1}$ and 
$\nu_b = 0.851^{+0.258}_{-0.137}$~d$^{-1}$, corresponding to 
characteristic timescales of approximately 3.2~d and 1.2~d, 
respectively. Sector~6 is best fit by the more complex model $M_3$ 
(bending power law B), with a bending frequency of 
$\nu_b = 1.194^{+0.023}_{-0.734}$~d$^{-1}$. The high-frequency 
spectral indices lie in the range $\alpha_1 \approx 2.1$--2.9 across 
all sectors, consistent with red-noise dominated variability as 
typically found in AGN and blazars \citep{2012A&A...544A..80G, 
2024MNRAS.527.2672S}. The preference for bending power laws over simple power 
laws has been observed in other TESS blazar studies \citep{2026ApJ...998..317K, 
2026MNRAS.545f1920T} and in long-term $\gamma$-ray PSD analyses of AGN 
\citep{2024MNRAS.527.2672S}, and is generally attributed to the presence of a 
characteristic timescale in the variability process.

A particularly noteworthy result emerges from comparing the PSD bending 
frequency with the detected QPO period in Sector~5. The best-fitting 
$M_2$ model for Sector~5 yields a bending frequency of 
$\nu_b = 0.308^{+0.103}_{-0.067}$~d$^{-1}$, corresponding to a 
characteristic timescale of $\sim$3.2~d. This is in close agreement 
with the QPO period of $P \approx 2.84$~d 
($f_\mathrm{QPO} = 0.352$~d$^{-1}$) detected in the same sector. The 
proximity of these two frequencies --- the PSD bending frequency and 
the quasi-periodic modulation --- is physically suggestive. In AGN and 
X-ray binary power spectra, a break in the PSD is commonly associated 
with a characteristic timescale of the underlying physical process 
driving the variability, such as particle cooling, light crossing, or 
the turnover of a damped oscillator \citep{2014ApJ...791...21F, sobolewska2014stochastic, 
2024MNRAS.527.2672S}. The coincidence of the PSD break timescale and the QPO 
period in Sector~5 suggests that these two phenomena may share a common 
physical origin rather than being independent features of the 
variability. Specifically, the PSD bending may reflect the same compact 
emitting region or instability timescale that produces the quasi-periodic 
modulation: the QPO manifests as a coherent oscillation driven by this 
process when the conditions are favorable, while the PSD break represents 
its imprint on the overall variability spectrum. A similar coincidence 
between the PSD bending frequency and the dominant variability timescale 
has been noted in analyses of blazar TESS light curves by 
\citet{2026MNRAS.545f1920T}, who found that Bartlett's periodogram and wavelet 
decomposition both place the dominant peak at frequencies consistent 
with the bending scale, and in $\gamma$-ray blazar PSDs where PSD breaks 
were interpreted as a signature of disc--jet coupling 
\citep{2024MNRAS.527.2672S}. In Sectors~6 and~32, where no QPO is detected, the 
PSD bending frequencies are substantially higher ($\sim$1.2~d$^{-1}$ 
and $\sim$0.85~d$^{-1}$, respectively), and no coherent oscillatory 
power accumulates near those scales, further supporting the view that 
the alignment of the break and QPO frequencies in Sector~5 is physically 
meaningful rather than coincidental. The bending frequency difference 
between sectors additionally confirms that the dominant variability 
timescale in PKS~0521$-$36 is not stationary but shifts between epochs, 
suggesting that the physical conditions in the compact emission region 
evolve on month-to-year timescales.

\subsection*{Flux Distribution}

The Anderson--Darling test firmly rejects both single Gaussian and 
single log-normal descriptions of the flux distribution in all three 
sectors, requiring two-component models. The double log-normal model 
provides the best or statistically equivalent description in each case. 
The positive skewness observed in all sectors (1.50, 0.78, and 0.68 
for Sectors~5, 6, and 32, respectively) and the presence of extended 
high-flux tails suggest sporadic flaring activity superposed on a 
quiescent baseline. A double log-normal flux distribution has been 
observed in other blazars and AGN at $\gamma$-ray, X-ray, and optical 
wavelengths \citep{Kushwaha_2016, 10.1093/mnras/stz3108, 2024MNRAS.527.2672S}, and has been 
associated with the presence of two distinct physical flux states, 
possibly corresponding to quiescent jet emission and active flaring 
episodes driven by particle acceleration events or shock propagation. 
A single log-normal distribution, which is typically linked to 
multiplicative accretion disk processes propagating into the jet 
\citep{Uttley_2001, 2024MNRAS.527.2672S}, is insufficient here, suggesting that 
the optical emission during these epochs involves at least two 
superposed processes. This finding is qualitatively consistent with 
the $\gamma$-ray flux analysis of PKS~0521$-$36 by \citet{2021ApJ...919...58Z}, 
who found that the long-term flux distribution is better described by 
a log-normal rather than a Gaussian model, and extends that result to 
the optical band at day-scale resolution.

\subsection*{QPO Detection and Physical Interpretation}

The most significant result of this study is the detection of a 
statistically robust quasi-periodic oscillation in the Sector~5 light 
curve. The LSP reveals a dominant peak at 
$f = 0.3524 \pm 0.0097$~d$^{-1}$, corresponding to a period of 
$P = 2.838 \pm 0.078$~d, exceeding the 99.99\% confidence level 
derived from $2 \times 10^4$ Monte Carlo simulations following the 
method of \citet{2013MNRAS.433..907E}. This detection is independently 
confirmed by the WWZ analysis, which yields a consistent peak at 
$f = 0.3522 \pm 0.0137$~d$^{-1}$ ($P = 2.839 \pm 0.110$~d) from the 
4-hour binned light curve. The WWZ time--frequency map demonstrates 
that the power is sustained throughout the 26.1-day baseline of 
Sector~5 rather than being confined to a brief interval, supporting 
a genuinely quasi-periodic rather than a purely transient fluctuation. 
The DRW-based null-hypothesis framework provides a further independent 
confirmation, with the LSP peak exceeding the $3\sigma$ confidence 
envelope constructed from $1.2 \times 10^4$ mock light curves, 
demonstrating that the signal cannot be attributed to red-noise 
fluctuations alone. The QPO spans approximately 9 coherent cycles 
within the sector baseline, well above the threshold of $\gtrsim 5$ 
cycles commonly required to distinguish a quasi-periodic feature from 
stochastic red-noise variability \citep{2016MNRAS.461.3145V}. No statistically 
significant periodic feature is found in Sectors~6 or 32, indicating 
that the oscillatory behavior is transient rather than a persistent 
characteristic of the source and is associated with the elevated 
activity state of Sector~5.

As discussed in the preceding subsection, the PSD bending frequency 
in Sector~5, $\nu_b \approx 0.308$~d$^{-1}$ (timescale $\sim$3.2~d), 
is in close agreement with the QPO frequency 
$f_\mathrm{QPO} \approx 0.352$~d$^{-1}$ (period $\sim$2.84~d). This 
alignment strengthens the physical significance of both detections: 
the PSD break and the QPO likely reflect the same underlying compact 
emission process, with the PSD break representing the characteristic 
frequency of the dominant variability driver and the QPO emerging as 
a coherent oscillatory mode of that same process during the active 
state of Sector~5.

This optical QPO detection enriches the emerging picture of 
multi-timescale quasi-periodic behavior in PKS~0521$-$36 established 
through $\gamma$-ray monitoring. Analyzing approximately 5.8 years of 
Fermi-LAT data bracketed between two major outbursts (MJD 56317--58447), 
\citet{2021ApJ...919...58Z} reported a $\gamma$-ray QPO at a period of 
$\sim$1.1~yr ($\sim$400~days) at approximately $5\sigma$ confidence, 
corroborated independently by the LSP, WWZ, REDFIT, and Gaussian 
process modeling. Extending the analysis to the full 15-year baseline, 
\citet{2023arXiv231212623S} identified three distinct QPO signatures at periods 
of $\sim$268, $\sim$295, and $\sim$806~days, with the 806-day signal 
appearing to correspond to the third harmonic of the 268-day 
oscillation. The optical QPO reported here at $P \approx 2.84$~d 
adds a previously uncharacterized short-timescale periodicity to this 
multiwavelength variability picture. The temporal baseline of Sector~5 
(MJD 58437--58462) falls immediately before the 2019 May outburst 
identified in the $\gamma$-ray band by \citet{2021ApJ...919...58Z}, placing the 
optical QPO during a period of rising activity in the source. This 
is consistent with a scenario in which the onset of a compact jet 
instability drives both the short-timescale optical oscillation and 
the subsequent major outburst, with the coherent periodic signature 
disappearing as the system enters the explosive flaring phase.

Short-timescale QPOs of a few days in AGN optical light curves are 
difficult to explain within the standard scenarios invoked for 
year-scale periodicities. Binary supermassive black hole systems 
\citep{1980Natur.287..307B} and large-scale jet precession driven by external 
torques \citep{2004ApJ...615L...5R} naturally produce characteristic periods 
orders of magnitude longer than observed here. The $\sim$2.84-day 
timescale instead points to compact physical processes operating 
within the inner accretion flow or at the base of the relativistic 
jet. We consider two physically motivated scenarios.

\paragraph{Disk-based hotspot or inner accretion flow oscillation.}
One possible interpretation is orbital motion of a non-axisymmetric 
structure --- such as a hotspot, spiral shock, or pulsation mode --- 
near the innermost stable circular orbit (ISCO) of the accretion disk 
\citep{1993ApJ...411..602C, 1993ApJ...406..420M, 2009ApJ...690..216G, 2008ApJ...679..182E}. 
Under this interpretation, the observed period corresponds to the 
orbital timescale at the ISCO, from which the black hole mass can be 
estimated using the standard relation \citep{2009ApJ...690..216G}:
\begin{equation}
    \frac{M_\mathrm{BH}}{M_\odot} = \frac{3.23 \times 10^4\, P}
    {(r^{3/2} + a)(1 + z)},
\end{equation}
where $P$ is the orbital period in seconds, $z = 0.055$ is the 
redshift of PKS~0521$-$36, $r$ is the ISCO radius in units of 
$GM/c^2$, and $a$ is the dimensionless spin parameter. For a 
Schwarzschild black hole ($r = 6$, $a = 0$) we obtain 
$M_\mathrm{BH} \approx 1.3 \times 10^{10}\,M_\odot$, which exceeds 
the typical range for AGN black holes and is therefore physically 
disfavored. For a maximally rotating Kerr black hole ($r = 1.2$, 
$a = 0.9982$) the inferred mass is 
$M_\mathrm{BH} \approx 2.5 \times 10^{9}\,M_\odot$, which, while 
at the high end, falls within the range spanned by FSRQs and 
powerful radio-loud AGN in the $\gamma$-ray QPO compilation of 
\citet{2023arXiv231212623S}. This scenario naturally accounts for the transient 
character of the QPO, since disk inhomogeneities are expected to 
dissipate over a finite number of orbital periods. However, in 
PKS~0521$-$36 the optical emission contains a significant non-thermal 
jet contribution \citep{2015MNRAS.450.3975D}, which would dilute any 
disk-hotspot signal and make a jet-based explanation comparably 
attractive. Furthermore, the coincidence of the PSD bending frequency 
with the QPO period, as discussed above, does not fit naturally into 
a pure disk-hotspot picture, since the PSD break in disk-dominated 
systems is generally associated with the viscous or thermal timescale 
at a characteristic disk radius rather than with a specific orbital 
frequency.

\paragraph{Magnetohydrodynamic kink instability in the relativistic jet.}
Although PKS~0521$-$36 is not a blazar, it hosts a well-established 
relativistic jet whose emission has been resolved across the radio, 
optical, and X-ray bands \citep{1979MNRAS.188..415D, 1991ApJ...369L..55M, 
2002MNRAS.335..142B, 2015MNRAS.450.3975D}. Crucially, \citet{2017MNRAS.470L.107J} 
directly detected helicoidal motion in the optical jet of 
PKS~0521$-$36 and interpreted it as evidence for a helical magnetic 
field structure or jet precession along the flow. The presence of a 
helical magnetic field is precisely the structural prerequisite for 
the development of current-driven magnetohydrodynamic kink 
instabilities \citep{2009ApJ...700..684M, 2020MNRAS.494.1817D}, making PKS~0521$-$36 
a physically motivated candidate for this mechanism independent of 
its non-blazar classification.

In a jet permeated by a helical or toroidal magnetic field, kink 
modes produce transverse displacements of the plasma column, 
distorting the local field geometry and triggering enhanced particle 
acceleration through magnetic reconnection events \citep{2020MNRAS.494.1817D}. 
Quasi-periodic compressions of the emitting region associated with 
the growth and propagation of a kink produce oscillatory flux 
enhancements on the characteristic kink growth timescale. In the 
observer's frame, this timescale is given by \citep{2020MNRAS.494.1817D}:
\begin{equation}
    T_\mathrm{obs} = \frac{R_\mathrm{KI}}{\langle v_\mathrm{tr} 
    \rangle\, \delta},
\end{equation}
where $R_\mathrm{KI}$ is the transverse size of the emitting region 
in the co-moving frame, $\langle v_\mathrm{tr} \rangle$ is the mean 
transverse propagation speed of the kink, and $\delta$ is the 
Doppler factor of the jet. The mildly beamed nature of 
PKS~0521$-$36 is not a limitation in this context; rather, the 
moderate Doppler factor naturally produces longer observed timescales 
than would be expected for a highly beamed blazar jet with 
$\delta \sim 20$--30, directly accounting for the day-scale period 
detected here rather than the sub-day timescales reported in some 
highly beamed blazars. Adopting $\langle v_\mathrm{tr} \rangle \approx 0.16c$ as found in relativistic MHD simulations 
\citep{2020MNRAS.494.1817D}, an emitting-region size 
$R_\mathrm{KI} = 10^{16}$--$10^{17}$~cm, and a Doppler factor in 
the range $\delta \approx 5$--10 consistent with the moderate beaming 
of PKS~0521$-$36 \citep{2016A&A...586A..70L, 2019A&A...627A.148A, 2015MNRAS.450.3975D}, the 
expected observer-frame timescale can be estimated explicitly. For 
$\langle v_\mathrm{tr} \rangle = 0.16c$, $R_\mathrm{KI} = 10^{16}$~cm, 
and $\delta = 5$, we obtain an observer-frame timescale of 
$T_\mathrm{obs} \approx 4.8~\mathrm{days}$.
While for $\delta = 10$ and the same emitting-region size the 
timescale reduces to $\sim$2.4~days. The range $\delta \approx 5$--10 
and $R_\mathrm{KI} = 10^{16}$--$10^{17}$~cm therefore spans an 
expected observer-frame timescale of approximately 2 to 20~days. The 
detected period of $\sim$2.84~d falls within this range, corresponding 
to a compact emitting region of size $R_\mathrm{KI} \sim 10^{16}$~cm 
for $\delta \approx 10$, consistent with the sub-parsec-scale 
$\gamma$-ray emission region inferred from the rapid variability 
timescales of $\sim$6--12~hr detected in this source 
\citep{2015MNRAS.450.3975D, 2019A&A...627A.148A, 2021ApJ...919...58Z}. We note that a Doppler 
factor of $\delta \approx 5$--10 lies at the moderate-to-upper end 
of estimates reported for PKS~0521$-$36 in the literature 
\citep{2016A&A...586A..70L, 2019A&A...627A.148A, 2015MNRAS.450.3975D}, but remains physically 
consistent with a mildly beamed jet viewed at a relatively small 
angle, and does not contradict the misaligned classification of the 
source.

This scenario also provides a natural and unified explanation for the 
coincidence between the PSD bending frequency and the QPO period in 
Sector~5. The best-fitting $M_2$ model for Sector~5 yields a bending 
frequency of $\nu_b = 0.308^{+0.103}_{-0.067}$~d$^{-1}$, 
corresponding to a characteristic timescale of $\sim$3.2~d, which is 
in close agreement with the QPO period of $P \approx 2.84$~d 
($f_\mathrm{QPO} = 0.352$~d$^{-1}$). The kink instability introduces 
a characteristic dynamical timescale into the jet that simultaneously 
sets the coherent oscillatory period of the QPO and the turnover 
frequency of the variability power spectrum. When the instability is 
active, power accumulates coherently at the kink growth frequency, 
producing both the PSD break and the QPO peak at the same 
characteristic scale. When the instability is suppressed --- as in 
Sectors~6 and~32, where the PSD bending frequencies are 
substantially higher ($\nu_b \approx 1.19$~d$^{-1}$ and 
$0.85$~d$^{-1}$, respectively) and no coherent oscillatory power 
accumulates --- the PSD is instead shaped by the shorter timescales 
of stochastic fluctuations in the jet plasma. This sector-to-sector 
shift in the PSD bending frequency, from $\sim$0.31~d$^{-1}$ in 
Sector~5 to $\sim$0.85$-$1.19~d$^{-1}$ in Sectors~32 and~6, 
confirms that the dominant variability timescale in the compact 
emission region is not stationary but evolves on month-to-year 
timescales, consistent with a jet whose physical conditions --- 
magnetic field strength, plasma injection rate, and bulk Lorentz 
factor --- change between epochs.

Kink instabilities are inherently transient: their development depends 
on the time-varying injection of magnetic energy and plasma into the 
jet flow, so that kink-driven modulations are expected to persist only 
for a limited number of cycles before the instability is suppressed 
or the associated emitting structure is advected downstream 
\citep{2020MNRAS.494.1817D}. This behavior is fully consistent with the 
non-detection of the QPO in Sectors~6 and~32 and with the sustained 
WWZ power across the full 26.1-day baseline of Sector~5. While 
observational precedents for kink-instability QPOs have so far been 
established predominantly in blazars --- including BL~Lacertae from 
TESS observations \citep{2022Natur.609..265J} and other TESS blazar studies 
\citep{2024MNRAS.527.9132T, 2026MNRAS.545f1920T, 2026arXiv260303984A} --- the underlying 
mechanism depends on the presence of a helical magnetic field and a 
relativistic jet, not on the jet orientation relative to the observer. 
PKS~0521$-$36, with its directly imaged helical jet structure 
\citep{2017MNRAS.470L.107J} and confirmed non-thermal jet emission 
across multiple wavelengths, represents one of the most physically 
motivated candidates for this mechanism outside the blazar class. 
While \citet{2021ApJ...919...58Z} previously reported the first $\gamma$-ray 
QPO in a mildly beamed jet from this same source, the optical 
detection reported here, interpreted within the kink-instability 
framework, constitutes to our knowledge the first such evidence at 
optical wavelengths for a non-blazar AGN with a directly imaged 
helical jet structure.

Other disk-based mechanisms --- normal modes of oscillation trapped 
in the innermost accretion flow by strong gravity \citep{1997ApJ...476..589P, 
2008ApJ...679..182E}, magnetorotational instability-driven turbulence 
\citep{2004ApJ...609L..63A}, and Lense--Thirring precession of a tilted 
inner disk \citep{1998ApJ...492L..59S} --- can in principle produce 
transient quasi-periodic modulations, but these processes are most 
relevant in sources where the optical emission is dominated by thermal 
accretion disk radiation. In PKS~0521$-$36, the non-thermal jet 
contribution is substantial across the optical band \citep{2015MNRAS.450.3975D}, 
making a jet-based mechanism the preferred physical explanation for 
the short-timescale optical QPO detected in Sector~5.

\subsection*{Multiwavelength Context and Outlook}

The transient character of the detected optical QPO --- present in 
Sector~5 but absent in Sectors~6 and~32 --- is a physically 
meaningful result that mirrors the behavior of the $\gamma$-ray QPO 
reported by \citet{2021ApJ...919...58Z}, which was confined to the interval 
between two major outbursts and was absent in the full 15-year 
Fermi-LAT dataset. Together, these observations suggest that 
PKS~0521$-$36 hosts a complex hierarchy of variability mechanisms 
operating simultaneously across a wide range of timescales: from 
compact jet instabilities producing day-scale optical QPOs, through 
intermediate-timescale oscillations on scales of months to years in 
the $\gamma$-ray band \citep{2021ApJ...919...58Z, 2023arXiv231212623S}, up to the major 
$\gamma$-ray outbursts. Whether the optical and $\gamma$-ray QPOs 
share a common physical origin --- for instance, if both are 
manifestations of jet instability at different spatial scales, or 
if they arise from unrelated processes at different locations in the 
source --- remains an open question.

Future high-cadence TESS observations of PKS~0521$-$36 in additional 
sectors, combined with simultaneous Fermi-LAT $\gamma$-ray monitoring, 
will be essential for establishing whether the $\sim$2.84-day optical 
QPO recurs during subsequent active phases, and for testing whether 
the PSD bending frequency and QPO period remain aligned during future 
outburst precursor states. The application of CARMA modeling and 
recurrence analysis to future multi-sector TESS datasets, following 
the approach of \citet{2026MNRAS.545f1920T}, will further constrain the 
stochastic and quasi-periodic components of the variability. Very 
long baseline interferometric imaging during optical QPO epochs could 
directly test the connection between the helicoidal jet structure 
detected by \citet{2017MNRAS.470L.107J} and the short-timescale optical 
flux modulations reported here, providing an observational link 
between the parsec-scale jet morphology and the compact variability 
processes in this remarkable non-blazar AGN.

\section{Acknowledgements}
ZS is supported by the Department of Science and Technology, Govt. of India, under the INSPIRE Faculty grant (DST/INSPIRE/04/2020/002319). SA and ZS express  gratitude to the Inter-University Centre for Astronomy and Astrophysics (IUCAA) in Pune, India, for the support and facilities provided.

\bibliographystyle{elsarticle-harv} 
\bibliography{sample631}

@article{Kushwaha_2016,
doi = {10.3847/2041-8205/822/1/L13},
url = {https://dx.doi.org/10.3847/2041-8205/822/1/L13},
year = {2016},
month = {apr},
publisher = {The American Astronomical Society},
volume = {822},
number = {1},
pages = {L13},
author = {Pankaj Kushwaha and Sunil Chandra and Ranjeev Misra and S. Sahayanathan and K. P. Singh and K. S. Baliyan},
title = {EVIDENCE FOR TWO LOGNORMAL STATES IN MULTI-WAVELENGTH FLUX VARIATION OF FSRQ PKS 1510-089},
journal = {The Astrophysical Journal Letters}
}

@article{uttley2005non,
  title={Non-linear X-ray variability in X-ray binaries and active galaxies},
  author={Uttley, Phil and McHardy, IM and Vaughan, S},
  journal={Monthly Notices of the Royal Astronomical Society},
  volume={359},
  number={1},
  pages={345--362},
  year={2005},
  publisher={Blackwell Science Ltd Oxford, UK}
}

@article{foster1996wavelets,
  title={Wavelets for period analysis of unevenly sampled time series},
  author={Foster, Grant},
  journal={Astronomical Journal v. 112, p. 1709-1729},
  volume={112},
  pages={1709--1729},
  year={1996}
}

@article{sobolewska2014stochastic,
  title={Stochastic modeling of the Fermi/LAT $\gamma$-ray blazar variability},
  author={Sobolewska, Malgorzata A and Siemiginowska, Aneta and Kelly, Brandon C and Nalewajko, Krzysztof},
  journal={The Astrophysical Journal},
  volume={786},
  number={2},
  pages={143},
  year={2014},
  publisher={IOP Publishing}
}

@article{moreno2019stochastic,
  title={Stochastic modeling handbook for optical AGN variability},
  author={Moreno, Jackeline and Vogeley, Michael S and Richards, Gordon T and Yu, Weixiang},
  journal={Publications of the Astronomical Society of the Pacific},
  volume={131},
  number={1000},
  pages={063001},
  year={2019},
  publisher={IOP Publishing}
}

@ARTICLE{2008MNRAS.385.1279B,
       author = {{Baluev}, R.~V.},
        title = "{Assessing the statistical significance of periodogram peaks}",
      journal = {\mnras},
         year = 2008,
        month = apr,
       volume = {385},
       number = {3},
        pages = {1279-1285},
          doi = {10.1111/j.1365-2966.2008.12689.x},
archivePrefix = {arXiv},
       eprint = {0711.0330},
 primaryClass = {astro-ph},
       adsurl = {https://ui.adsabs.harvard.edu/abs/2008MNRAS.385.1279B}
}

@ARTICLE{2004ApJ...615L...5R,
       author = {{Rieger}, Frank M.},
        title = "{On the Geometrical Origin of Periodicity in Blazar-type Sources}",
      journal = {\apjl},
         year = 2004,
        month = nov,
       volume = {615},
       number = {1},
        pages = {L5-L8},
          doi = {10.1086/426018},
archivePrefix = {arXiv},
       eprint = {astro-ph/0410188},
 primaryClass = {astro-ph},
       adsurl = {https://ui.adsabs.harvard.edu/abs/2004ApJ...615L...5R}
}

@article{zxgv-fzv5,
  title = {Indication for dual periodic signatures in PKS 0805-07 from multitechnique time series analysis},
  author = {Akbar, Sikandar and Shah, Zahir and Misra, Ranjeev and Boked, Sajad and Iqbal, Naseer},
  journal = {Phys. Rev. D},
  volume = {112},
  issue = {6},
  pages = {063061},
  numpages = {20},
  year = {2025},
  month = {Sep},
  publisher = {American Physical Society},
  doi = {10.1103/zxgv-fzv5},
  url = {https://link.aps.org/doi/10.1103/zxgv-fzv5}
}

@article{TANTRY2025100372,
title = {Study of multi-wavelength variability, emission mechanism and quasi-periodic oscillation for transition blazar S5 1803+784},
journal = {Journal of High Energy Astrophysics},
volume = {47},
pages = {100372},
year = {2025},
issn = {2214-4048},
doi = {https://doi.org/10.1016/j.jheap.2025.100372},
url = {https://www.sciencedirect.com/science/article/pii/S2214404825000539},
author = {Javaid Tantry and Ajay Sharma and Zahir Shah and Naseer Iqbal and Debanjan Bose}
}

@article{burke2021characteristic,
  title={A characteristic optical variability time scale in astrophysical accretion disks},
  author={Burke, Colin J and Shen, Yue and Blaes, Omer and Gammie, Charles F and Horne, Keith and Jiang, Yan-Fei and Liu, Xin and McHardy, Ian M and Morgan, Christopher W and Scaringi, Simone and others},
  journal={Science},
  volume={373},
  number={6556},
  pages={789--792},
  year={2021},
  publisher={American Association for the Advancement of Science}
}

@article{zhang2022characterizing,
  title={Characterizing the $\gamma$-ray variability of active galactic nuclei with the stochastic process method},
  author={Zhang, Haiyun and Yan, Dahai and Zhang, Li},
  journal={The Astrophysical Journal},
  volume={930},
  number={2},
  pages={157},
  year={2022},
  publisher={IOP Publishing}
}

@article{zhang2023gaussian,
  title={Gaussian Process Modeling Blazar Multiwavelength Variability: Indirectly Resolving Jet Structure},
  author={Zhang, Haiyun and Yan, Dahai and Zhang, Li},
  journal={The Astrophysical Journal},
  volume={944},
  number={1},
  pages={103},
  year={2023},
  publisher={IOP Publishing}
}

@article{sharma2024microquasars,
  title={Microquasars to AGNs: An uniform Jet variability},
  author={Sharma, Ajay and Prince, Raj and Bose, Debanjan},
  journal={arXiv preprint arXiv:2410.06653},
  year={2024}
}

@ARTICLE{1995PASP..107..803U,
       author = {{Urry}, C. Megan and {Padovani}, Paolo},
        title = "{Unified Schemes for Radio-Loud Active Galactic Nuclei}",
      journal = {\pasp},
         year = 1995,
        month = sep,
       volume = {107},
        pages = {803},
          doi = {10.1086/133630},
archivePrefix = {arXiv},
       eprint = {astro-ph/9506063},
 primaryClass = {astro-ph},
       adsurl = {https://ui.adsabs.harvard.edu/abs/1995PASP..107..803U}
}

@article{Nazir_2026,
doi = {10.3847/1538-4357/ae3b3d},
url = {https://doi.org/10.3847/1538-4357/ae3b3d},
year = {2026},
month = {feb},
publisher = {The American Astronomical Society},
volume = {998},
number = {2},
pages = {227},
author = {Nazir, Zeeshan and Akbar, Sikandar and Shah, Zahir and Dar, Athar A. and Malik, Zahoor},
title = {Broadband Variability Analysis of the Flat-spectrum Radio Quasar PKS 0402-362 with Indications of Quasi-periodic Modulation},
journal = {The Astrophysical Journal}
}

@ARTICLE{1998ApJ...492L..59S,
       author = {{Stella}, Luigi and {Vietri}, Mario},
        title = "{Lense-Thirring Precession and Quasi-periodic Oscillations in Low-Mass X-Ray Binaries}",
      journal = {\apjl},
         year = 1998,
        month = jan,
       volume = {492},
       number = {1},
        pages = {L59-L62},
          doi = {10.1086/311075},
archivePrefix = {arXiv},
       eprint = {astro-ph/9709085},
 primaryClass = {astro-ph},
       adsurl = {https://ui.adsabs.harvard.edu/abs/1998ApJ...492L..59S}
}

@ARTICLE{1993ApJ...411..602C,
       author = {{Chakrabarti}, Sandip K. and {Wiita}, Paul J.},
        title = "{Spiral Shocks in Accretion Disks As a Contributor to Variability in Active Galactic Nuclei}",
      journal = {\apj},
         year = 1993,
        month = jul,
       volume = {411},
        pages = {602},
          doi = {10.1086/172862},
       adsurl = {https://ui.adsabs.harvard.edu/abs/1993ApJ...411..602C}
}

@ARTICLE{1993ApJ...406..420M,
       author = {{Mangalam}, Arun V. and {Wiita}, Paul J.},
        title = "{Accretion Disk Models for Optical and Ultraviolet Microvariability in Active Galactic Nuclei}",
      journal = {\apj},
         year = 1993,
        month = apr,
       volume = {406},
        pages = {420},
          doi = {10.1086/172453},
       adsurl = {https://ui.adsabs.harvard.edu/abs/1993ApJ...406..420M}
}

@ARTICLE{2008ApJ...679..182E,
       author = {{Espaillat}, C. and {Bregman}, J. and {Hughes}, P. and {Lloyd-Davies}, E.},
        title = "{Wavelet Analysis of AGN X-Ray Time Series: A QPO in 3C 273?}",
      journal = {\apj},
         year = 2008,
        month = may,
       volume = {679},
       number = {1},
        pages = {182-193},
          doi = {10.1086/587023},
archivePrefix = {arXiv},
       eprint = {0805.4342},
 primaryClass = {astro-ph},
       adsurl = {https://ui.adsabs.harvard.edu/abs/2008ApJ...679..182E}
}

@ARTICLE{2009ApJ...690..216G,
       author = {{Gupta}, Alok C. and {Srivastava}, A.~K. and {Wiita}, Paul J.},
        title = "{Periodic Oscillations in the Intra-Day Optical Light Curves of the Blazar S5 0716+714}",
      journal = {\apj},
         year = 2009,
        month = jan,
       volume = {690},
       number = {1},
        pages = {216-223},
          doi = {10.1088/0004-637X/690/1/216},
archivePrefix = {arXiv},
       eprint = {0808.3630},
 primaryClass = {astro-ph},
       adsurl = {https://ui.adsabs.harvard.edu/abs/2009ApJ...690..216G}
}

@ARTICLE{2020MNRAS.494.1817D,
       author = {{Dong}, Lingyi and {Zhang}, Haocheng and {Giannios}, Dimitrios},
        title = "{Kink instabilities in relativistic jets can drive quasi-periodic radiation signatures}",
      journal = {\mnras},
         year = 2020,
        month = may,
       volume = {494},
       number = {2},
        pages = {1817-1825},
          doi = {10.1093/mnras/staa773},
archivePrefix = {arXiv},
       eprint = {2003.07765},
 primaryClass = {astro-ph.HE},
       adsurl = {https://ui.adsabs.harvard.edu/abs/2020MNRAS.494.1817D}
}

@ARTICLE{2009ApJ...700..684M,
       author = {{Mizuno}, Yosuke and {Lyubarsky}, Yuri and {Nishikawa}, Ken-Ichi and {Hardee}, Philip E.},
        title = "{Three-Dimensional Relativistic Magnetohydrodynamic Simulations of Current-Driven Instability. I. Instability of a Static Column}",
      journal = {\apj},
         year = 2009,
        month = jul,
       volume = {700},
       number = {1},
        pages = {684-693},
          doi = {10.1088/0004-637X/700/1/684},
archivePrefix = {arXiv},
       eprint = {0903.5358},
 primaryClass = {astro-ph.HE},
       adsurl = {https://ui.adsabs.harvard.edu/abs/2009ApJ...700..684M}
}

@ARTICLE{2022Natur.609..265J,
       author = {{Jorstad}, S.~G. and {Marscher}, A.~P. and {Raiteri}, C.~M. and {Villata}, M. and {Weaver}, Z.~R. and {Zhang}, H. and {Dong}, L. and {G{\'o}mez}, J.~L. and {Perel}, M.~V. and {Savchenko}, S.~S. and {Larionov}, V.~M. and {Carosati}, D. and {Chen}, W.~P. and {Kurtanidze}, O.~M. and {Marchini}, A. and {Matsumoto}, K. and {Mortari}, F. and {Aceti}, P. and {Acosta-Pulido}, J.~A. and {Andreeva}, T. and {Apolonio}, G. and {Arena}, C. and {Arkharov}, A. and {Bachev}, R. and {Banfi}, M. and {Bonnoli}, G. and {Borman}, G.~A. and {Bozhilov}, V. and {Carnerero}, M.~I. and {Damljanovic}, G. and {Ehgamberdiev}, S.~A. and {Els{\"a}sser}, D. and {Frasca}, A. and {Gabellini}, D. and {Grishina}, T.~S. and {Gupta}, A.~C. and {Hagen-Thorn}, V.~A. and {Hallum}, M.~K. and {Hart}, M. and {Hasuda}, K. and {Hemrich}, F. and {Hsiao}, H.~Y. and {Ibryamov}, S. and {Irsmambetova}, T.~R. and {Ivanov}, D.~V. and {Joner}, M.~D. and {Kimeridze}, G.~N. and {Klimanov}, S.~A. and {Kn{\"o}tt}, J. and {Kopatskaya}, E.~N. and {Kurtanidze}, S.~O. and {Kurtenkov}, A. and {Kuutma}, T. and {Larionova}, E.~G. and {Leonini}, S. and {Lin}, H.~C. and {Lorey}, C. and {Mannheim}, K. and {Marino}, G. and {Minev}, M. and {Mirzaqulov}, D.~O. and {Morozova}, D.~A. and {Nikiforova}, A.~A. and {Nikolashvili}, M.~G. and {Ovcharov}, E. and {Papini}, R. and {Pursimo}, T. and {Rahimov}, I. and {Reinhart}, D. and {Sakamoto}, T. and {Salvaggio}, F. and {Semkov}, E. and {Shakhovskoy}, D.~N. and {Sigua}, L.~A. and {Steineke}, R. and {Stojanovic}, M. and {Strigachev}, A. and {Troitskaya}, Y.~V. and {Troitskiy}, I.~S. and {Tsai}, A. and {Valcheva}, A. and {Vasilyev}, A.~A. and {Vince}, O. and {Waller}, L. and {Zaharieva}, E. and {Chatterjee}, R.},
        title = "{Rapid quasi-periodic oscillations in the relativistic jet of BL Lacertae}",
      journal = {Nature},
         year = 2022,
        month = sep,
       volume = {609},
       number = {7926},
        pages = {265-268},
          doi = {10.1038/s41586-022-05038-9},
       adsurl = {https://ui.adsabs.harvard.edu/abs/2022Natur.609..265J}
}

@ARTICLE{2004ApJ...609L..63A,
       author = {{Abramowicz}, Marek A. and {Klu{\'z}niak}, W{\l}odek and {McClintock}, Jeffrey E. and {Remillard}, Ronald A.},
        title = "{The Importance of Discovering a 3:2 Twin-Peak Quasi-periodic Oscillation in an Ultraluminous X-Ray Source, or How to Solve the Puzzle of Intermediate-Mass Black Holes}",
      journal = {\apjl},
         year = 2004,
        month = jul,
       volume = {609},
       number = {2},
        pages = {L63-L65},
          doi = {10.1086/422810},
archivePrefix = {arXiv},
       eprint = {astro-ph/0402012},
 primaryClass = {astro-ph},
       adsurl = {https://ui.adsabs.harvard.edu/abs/2004ApJ...609L..63A}
}

@ARTICLE{1997ApJ...476..589P,
       author = {{Perez}, Christopher A. and {Silbergleit}, Alexander S. and {Wagoner}, Robert V. and {Lehr}, Dana E.},
        title = "{Relativistic Diskoseismology. I. Analytical Results for ``Gravity Modes''}",
      journal = {\apj},
         year = 1997,
        month = feb,
       volume = {476},
       number = {2},
        pages = {589-604},
          doi = {10.1086/303658},
archivePrefix = {arXiv},
       eprint = {astro-ph/9601146},
 primaryClass = {astro-ph},
       adsurl = {https://ui.adsabs.harvard.edu/abs/1997ApJ...476..589P}
}

@ARTICLE{2016MNRAS.461.3145V,
       author = {{Vaughan}, S. and {Uttley}, P. and {Markowitz}, A.~G. and {Huppenkothen}, D. and {Middleton}, M.~J. and {Alston}, W.~N. and {Scargle}, J.~D. and {Farr}, W.~M.},
        title = "{False periodicities in quasar time-domain surveys}",
      journal = {\mnras},
         year = 2016,
        month = sep,
       volume = {461},
       number = {3},
        pages = {3145-3152},
          doi = {10.1093/mnras/stw1412},
archivePrefix = {arXiv},
       eprint = {1606.02620},
 primaryClass = {astro-ph.IM},
       adsurl = {https://ui.adsabs.harvard.edu/abs/2016MNRAS.461.3145V}
}

@ARTICLE{2015JATIS...1a4003R,
       author = {{Ricker}, George R. and {Winn}, Joshua N. and {Vanderspek}, Roland and {Latham}, David W. and {Bakos}, G{\'a}sp{\'a}r {\'A}. and {Bean}, Jacob L. and {Berta-Thompson}, Zachory K. and {Brown}, Timothy M. and {Buchhave}, Lars and {Butler}, Nathaniel R. and {Butler}, R. Paul and {Chaplin}, William J. and {Charbonneau}, David and {Christensen-Dalsgaard}, J{\o}rgen and {Clampin}, Mark and {Deming}, Drake and {Doty}, John and {De Lee}, Nathan and {Dressing}, Courtney and {Dunham}, Edward W. and {Endl}, Michael and {Fressin}, Francois and {Ge}, Jian and {Henning}, Thomas and {Holman}, Matthew J. and {Howard}, Andrew W. and {Ida}, Shigeru and {Jenkins}, Jon M. and {Jernigan}, Garrett and {Johnson}, John Asher and {Kaltenegger}, Lisa and {Kawai}, Nobuyuki and {Kjeldsen}, Hans and {Laughlin}, Gregory and {Levine}, Alan M. and {Lin}, Douglas and {Lissauer}, Jack J. and {MacQueen}, Phillip and {Marcy}, Geoffrey and {McCullough}, Peter R. and {Morton}, Timothy D. and {Narita}, Norio and {Paegert}, Martin and {Palle}, Enric and {Pepe}, Francesco and {Pepper}, Joshua and {Quirrenbach}, Andreas and {Rinehart}, Stephen A. and {Sasselov}, Dimitar and {Sato}, Bun'ei and {Seager}, Sara and {Sozzetti}, Alessandro and {Stassun}, Keivan G. and {Sullivan}, Peter and {Szentgyorgyi}, Andrew and {Torres}, Guillermo and {Udry}, Stephane and {Villasenor}, Joel},
        title = "{Transiting Exoplanet Survey Satellite (TESS)}",
      journal = {Journal of Astronomical Telescopes, Instruments, and Systems},
         year = 2015,
        month = jan,
       volume = {1},
          eid = {014003},
        pages = {014003},
          doi = {10.1117/1.JATIS.1.1.014003},
       adsurl = {https://ui.adsabs.harvard.edu/abs/2015JATIS...1a4003R}
}

@article{Smith_2023,
doi = {10.3847/1538-4357/acff5c},
url = {https://doi.org/10.3847/1538-4357/acff5c},
year = {2023},
month = {nov},
publisher = {The American Astronomical Society},
volume = {958},
number = {2},
pages = {188},
author = {Smith, Krista Lynne and Sartori, Lia F.},
title = {The Rapid Optical Variability of the Nearby Radio-loud AGN Pictor A: Introducing the Quaver Pipeline for AGN Science with TESS},
journal = {The Astrophysical Journal}
}

@ARTICLE{2024MNRAS.527.9132T,
       author = {{Tripathi}, Ashutosh and {Smith}, Krista Lynne and {Wiita}, Paul J. and {Wagoner}, Robert V.},
        title = "{Optical quasi-periodic oscillations in the TESS light curves of three blazars}",
      journal = {\mnras},
         year = 2024,
        month = jan,
       volume = {527},
       number = {3},
        pages = {9132-9144},
          doi = {10.1093/mnras/stad3744},
archivePrefix = {arXiv},
       eprint = {2312.14144},
 primaryClass = {astro-ph.HE},
       adsurl = {https://ui.adsabs.harvard.edu/abs/2024MNRAS.527.9132T}
}

@ARTICLE{2026arXiv260120471A,
       author = {{Akbar}, Sikandar},
        title = "{A multi-technique search for year-scale $\gamma$-ray quasi-periodic modulation in the high-redshift FSRQ PKS\raisebox{-0.5ex}\textasciitilde2052$-$47}",
      journal = {arXiv e-prints},
         year = 2026,
        month = jan,
          eid = {arXiv:2601.20471},
        pages = {arXiv:2601.20471},
          doi = {10.48550/arXiv.2601.20471},
archivePrefix = {arXiv},
       eprint = {2601.20471},
 primaryClass = {astro-ph.HE},
       adsurl = {https://ui.adsabs.harvard.edu/abs/2026arXiv260120471A}
}

@ARTICLE{2017A&ARv..25....2P,
       author = {{Padovani}, P. and {Alexander}, D.~M. and {Assef}, R.~J. and {De Marco}, B. and {Giommi}, P. and {Hickox}, R.~C. and {Richards}, G.~T. and {Smol{\v{c}}i{\'c}}, V. and {Hatziminaoglou}, E. and {Mainieri}, V. and {Salvato}, M.},
        title = "{Active galactic nuclei: what's in a name?}",
      journal = {The Astronomy and Astrophysics Review},
         year = 2017,
        month = aug,
       volume = {25},
       number = {1},
          eid = {2},
        pages = {2},
          doi = {10.1007/s00159-017-0102-9},
archivePrefix = {arXiv},
       eprint = {1707.07134},
 primaryClass = {astro-ph.GA},
       adsurl = {https://ui.adsabs.harvard.edu/abs/2017A&ARv..25....2P}
}

@ARTICLE{2016ARA&A..54..725M,
       author = {{Madejski}, Grzegorz (Greg) and {Sikora}, Marek},
        title = "{Gamma-Ray Observations of Active Galactic Nuclei}",
      journal = {\araa},
         year = 2016,
        month = sep,
       volume = {54},
        pages = {725-760},
          doi = {10.1146/annurev-astro-081913-040044},
       adsurl = {https://ui.adsabs.harvard.edu/abs/2016ARA&A..54..725M}
}

@ARTICLE{2019ARA&A..57..467B,
       author = {{Blandford}, Roger and {Meier}, David and {Readhead}, Anthony},
        title = "{Relativistic Jets from Active Galactic Nuclei}",
      journal = {\araa},
         year = 2019,
        month = aug,
       volume = {57},
        pages = {467-509},
          doi = {10.1146/annurev-astro-081817-051948},
archivePrefix = {arXiv},
       eprint = {1812.06025},
 primaryClass = {astro-ph.HE},
       adsurl = {https://ui.adsabs.harvard.edu/abs/2019ARA&A..57..467B}
}

@ARTICLE{2019Galax...7...20B,
       author = {{B{\"o}ttcher}, Markus},
        title = "{Progress in Multi-Wavelength and Multi-Messenger Observations of Blazars and Theoretical Challenges}",
      journal = {Galaxies},
         year = 2019,
        month = jan,
       volume = {7},
       number = {1},
          eid = {20},
        pages = {20},
          doi = {10.3390/galaxies7010020},
archivePrefix = {arXiv},
       eprint = {1901.04178},
 primaryClass = {astro-ph.HE},
       adsurl = {https://ui.adsabs.harvard.edu/abs/2019Galax...7...20B}
}

@ARTICLE{1986ApJ...302..296K,
       author = {{Keel}, W.~C.},
        title = "{The Jet of PKS 0521-36: an Aging Counterpart of M87?}",
      journal = {\apj},
         year = 1986,
        month = mar,
       volume = {302},
        pages = {296},
          doi = {10.1086/163991},
       adsurl = {https://ui.adsabs.harvard.edu/abs/1986ApJ...302..296K}
}

@ARTICLE{1995A&A...303..730S,
       author = {{Scarpa}, R. and {Falomo}, R. and {Pian}, E.},
        title = "{A study of emission lines variability of the active galaxy PKS 0521-365.}",
      journal = {\aap},
         year = 1995,
        month = nov,
       volume = {303},
        pages = {730},
       adsurl = {https://ui.adsabs.harvard.edu/abs/1995A&A...303..730S}
}

@ARTICLE{2015MNRAS.450.3975D,
       author = {{D'Ammando}, F. and {Orienti}, M. and {Tavecchio}, F. and {Ghisellini}, G. and {Torresi}, E. and {Giroletti}, M. and {Raiteri}, C.~M. and {Grandi}, P. and {Aller}, M. and {Aller}, H. and {Gurwell}, M.~A. and {Malaguti}, G. and {Pian}, E. and {Tosti}, G.},
        title = "{Unveiling the nature of the {\ensuremath{\gamma}}-ray emitting active galactic nucleus PKS 0521-36}",
      journal = {\mnras},
         year = 2015,
        month = jul,
       volume = {450},
       number = {4},
        pages = {3975-3990},
          doi = {10.1093/mnras/stv909},
archivePrefix = {arXiv},
       eprint = {1504.05595},
 primaryClass = {astro-ph.HE},
       adsurl = {https://ui.adsabs.harvard.edu/abs/2015MNRAS.450.3975D}
}

@ARTICLE{2020ApJS..247...33A,
       author = {{Abdollahi}, S. and {Acero}, F. and {Ackermann}, M. and {Ajello}, M. and {Atwood}, W.~B. and {Axelsson}, M. and {Baldini}, L. and {Ballet}, J. and {Barbiellini}, G. and {Bastieri}, D. and {Becerra Gonzalez}, J. and {Bellazzini}, R. and {Berretta}, A. and {Bissaldi}, E. and {Blandford}, R.~D. and {Bloom}, E.~D. and {Bonino}, R. and {Bottacini}, E. and {Brandt}, T.~J. and {Bregeon}, J. and {Bruel}, P. and {Buehler}, R. and {Burnett}, T.~H. and {Buson}, S. and {Cameron}, R.~A. and {Caputo}, R. and {Caraveo}, P.~A. and {Casandjian}, J.~M. and {Castro}, D. and {Cavazzuti}, E. and {Charles}, E. and {Chaty}, S. and {Chen}, S. and {Cheung}, C.~C. and {Chiaro}, G. and {Ciprini}, S. and {Cohen-Tanugi}, J. and {Cominsky}, L.~R. and {Coronado-Bl{\'a}zquez}, J. and {Costantin}, D. and {Cuoco}, A. and {Cutini}, S. and {D'Ammando}, F. and {DeKlotz}, M. and {de la Torre Luque}, P. and {de Palma}, F. and {Desai}, A. and {Digel}, S.~W. and {Di Lalla}, N. and {Di Mauro}, M. and {Di Venere}, L. and {Dom{\'\i}nguez}, A. and {Dumora}, D. and {Fana Dirirsa}, F. and {Fegan}, S.~J. and {Ferrara}, E.~C. and {Franckowiak}, A. and {Fukazawa}, Y. and {Funk}, S. and {Fusco}, P. and {Gargano}, F. and {Gasparrini}, D. and {Giglietto}, N. and {Giommi}, P. and {Giordano}, F. and {Giroletti}, M. and {Glanzman}, T. and {Green}, D. and {Grenier}, I.~A. and {Griffin}, S. and {Grondin}, M.-H. and {Grove}, J.~E. and {Guiriec}, S. and {Harding}, A.~K. and {Hayashi}, K. and {Hays}, E. and {Hewitt}, J.~W. and {Horan}, D. and {J{\'o}hannesson}, G. and {Johnson}, T.~J. and {Kamae}, T. and {Kerr}, M. and {Kocevski}, D. and {Kovac'evic'}, M. and {Kuss}, M. and {Landriu}, D. and {Larsson}, S. and {Latronico}, L. and {Lemoine-Goumard}, M. and {Li}, J. and {Liodakis}, I. and {Longo}, F. and {Loparco}, F. and {Lott}, B. and {Lovellette}, M.~N. and {Lubrano}, P. and {Madejski}, G.~M. and {Maldera}, S. and {Malyshev}, D. and {Manfreda}, A. and {Marchesini}, E.~J. and {Marcotulli}, L. and {Mart{\'\i}-Devesa}, G. and {Martin}, P. and {Massaro}, F. and {Mazziotta}, M.~N. and {McEnery}, J.~E. and {Mereu}, I. and {Meyer}, M. and {Michelson}, P.~F. and {Mirabal}, N. and {Mizuno}, T. and {Monzani}, M.~E. and {Morselli}, A. and {Moskalenko}, I.~V. and {Negro}, M. and {Nuss}, E. and {Ojha}, R. and {Omodei}, N. and {Orienti}, M. and {Orlando}, E. and {Ormes}, J.~F. and {Palatiello}, M. and {Paliya}, V.~S. and {Paneque}, D. and {Pei}, Z. and {Pe{\~n}a-Herazo}, H. and {Perkins}, J.~S. and {Persic}, M. and {Pesce-Rollins}, M. and {Petrosian}, V. and {Petrov}, L. and {Piron}, F. and {Poon}, H. and {Porter}, T.~A. and {Principe}, G. and {Rain{\`o}}, S. and {Rando}, R. and {Razzano}, M. and {Razzaque}, S. and {Reimer}, A. and {Reimer}, O. and {Remy}, Q. and {Reposeur}, T. and {Romani}, R.~W. and {Saz Parkinson}, P.~M. and {Schinzel}, F.~K. and {Serini}, D. and {Sgr{\`o}}, C. and {Siskind}, E.~J. and {Smith}, D.~A. and {Spandre}, G. and {Spinelli}, P. and {Strong}, A.~W. and {Suson}, D.~J. and {Tajima}, H. and {Takahashi}, M.~N. and {Tak}, D. and {Thayer}, J.~B. and {Thompson}, D.~J. and {Tibaldo}, L. and {Torres}, D.~F. and {Torresi}, E. and {Valverde}, J. and {Van Klaveren}, B. and {van Zyl}, P. and {Wood}, K. and {Yassine}, M. and {Zaharijas}, G.},
        title = "{Fermi Large Area Telescope Fourth Source Catalog}",
      journal = {\apjs},
         year = 2020,
        month = mar,
       volume = {247},
       number = {1},
          eid = {33},
        pages = {33},
          doi = {10.3847/1538-4365/ab6bcb},
archivePrefix = {arXiv},
       eprint = {1902.10045},
 primaryClass = {astro-ph.HE},
       adsurl = {https://ui.adsabs.harvard.edu/abs/2020ApJS..247...33A}
}

@ARTICLE{2019A&A...627A.148A,
       author = {{Angioni}, R. and {Ros}, E. and {Kadler}, M. and {Ojha}, R. and {M{\"u}ller}, C. and {Edwards}, P.~G. and {Burd}, P.~R. and {Carpenter}, B. and {Dutka}, M.~S. and {Gulyaev}, S. and {Hase}, H. and {Horiuchi}, S. and {Krau{\ss}}, F. and {Lovell}, J.~E.~J. and {Natusch}, T. and {Phillips}, C. and {Pl{\"o}tz}, C. and {Quick}, J.~F.~H. and {R{\"o}sch}, F. and {Schulz}, R. and {Stevens}, J. and {Tzioumis}, A.~K. and {Weston}, S. and {Wilms}, J. and {Zensus}, J.~A.},
        title = "{Gamma-ray emission in radio galaxies under the VLBI scope. I. Parsec-scale jet kinematics and high-energy properties of {\ensuremath{\gamma}}-ray-detected TANAMI radio galaxies}",
      journal = {\aap},
         year = 2019,
        month = jul,
       volume = {627},
          eid = {A148},
        pages = {A148},
          doi = {10.1051/0004-6361/201935697},
archivePrefix = {arXiv},
       eprint = {1906.08342},
 primaryClass = {astro-ph.HE},
       adsurl = {https://ui.adsabs.harvard.edu/abs/2019A&A...627A.148A}
}

@ARTICLE{1999ApJ...526..643S,
       author = {{Scarpa}, Riccardo and {Urry}, C. Megan and {Falomo}, Renato and {Treves}, Aldo},
        title = "{Hubble Space Telescope Observations of the Optical Jets of PKS 0521-365, 3C 371, and PKS 2201+044}",
      journal = {\apj},
         year = 1999,
        month = dec,
       volume = {526},
       number = {2},
        pages = {643-648},
          doi = {10.1086/308041},
archivePrefix = {arXiv},
       eprint = {astro-ph/9906462},
 primaryClass = {astro-ph},
       adsurl = {https://ui.adsabs.harvard.edu/abs/1999ApJ...526..643S}
}

@ARTICLE{1979MNRAS.188..415D,
       author = {{Danziger}, I.~J. and {Fosbury}, R.~A.~E. and {Goss}, W.~M. and {Ekers}, R.~D.},
        title = "{The radio and optical properties of the BL Lac object PKS 0521-36.}",
      journal = {\mnras},
         year = 1979,
        month = aug,
       volume = {188},
        pages = {415-419},
          doi = {10.1093/mnras/188.2.415},
       adsurl = {https://ui.adsabs.harvard.edu/abs/1979MNRAS.188..415D}
}

@ARTICLE{1991ApJ...369L..55M,
       author = {{Macchetto}, F. and {Albrecht}, R. and {Barbieri}, C. and {Blades}, J.~C. and {Boksenberg}, A. and {Crane}, P. and {Deharveng}, J.~M. and {Disney}, M.~J. and {Jakobsen}, P. and {Kamperman}, T.~M. and {King}, I.~R. and {Mackay}, C.~D. and {Paresce}, F. and {Weigelt}, G. and {Baxter}, D. and {Greenfield}, P. and {Jedrzejewski}, R. and {Nota}, A. and {Sparks}, W.~B.},
        title = "{First Results from the Faint Object Camera: Observations of PKS 0521-36}",
      journal = {\apjl},
         year = 1991,
        month = mar,
       volume = {369},
        pages = {L55},
          doi = {10.1086/185957},
       adsurl = {https://ui.adsabs.harvard.edu/abs/1991ApJ...369L..55M}
}

@ARTICLE{2002MNRAS.335..142B,
       author = {{Birkinshaw}, M. and {Worrall}, D.~M. and {Hardcastle}, M.~J.},
        title = "{The X-ray jet and halo of PKS 0521-365}",
      journal = {\mnras},
         year = 2002,
        month = sep,
       volume = {335},
       number = {1},
        pages = {142-150},
          doi = {10.1046/j.1365-8711.2002.05615.x},
archivePrefix = {arXiv},
       eprint = {astro-ph/0204509},
 primaryClass = {astro-ph},
       adsurl = {https://ui.adsabs.harvard.edu/abs/2002MNRAS.335..142B}
}

@INPROCEEDINGS{2017ques.workE..16L,
       author = {{Liuzzo}, Elisabetta and {Paladino}, R. and {Galluzzi}, V.},
        title = "{Polarized Emission In The Mm Band Of Pks0521-365: Alma Observations}",
    booktitle = {Submm/mm/cm QUESO Workshop 2017 (QUESO2017)},
         year = 2017,
        month = oct,
          eid = {16},
        pages = {16},
          doi = {10.5281/zenodo.1038079},
       adsurl = {https://ui.adsabs.harvard.edu/abs/2017ques.workE..16L}
}

@ARTICLE{2009A&A...501..907F,
       author = {{Falomo}, R. and {Pian}, E. and {Treves}, A. and {Giovannini}, G. and {Venturi}, T. and {Moretti}, A. and {Arcidiacono}, C. and {Farinato}, J. and {Ragazzoni}, R. and {Diolaiti}, E. and {Lombini}, M. and {Tavecchio}, F. and {Brast}, R. and {Donaldson}, R. and {Kolb}, J. and {Marchetti}, E. and {Tordo}, S.},
        title = "{The jet of the BL Lacertae object PKS 0521-365 in the near-IR: MAD adaptive optics observations}",
      journal = {\aap},
         year = 2009,
        month = jul,
       volume = {501},
       number = {3},
        pages = {907-914},
          doi = {10.1051/0004-6361/200912077},
archivePrefix = {arXiv},
       eprint = {0906.1069},
 primaryClass = {astro-ph.CO},
       adsurl = {https://ui.adsabs.harvard.edu/abs/2009A&A...501..907F}
}

@ARTICLE{2016A&A...586A..70L,
       author = {{Leon}, S. and {Cortes}, P.~C. and {Guerard}, M. and {Villard}, E. and {Hidayat}, T. and {Oca{\~n}a Flaquer}, B. and {Vila-Vilaro}, B.},
        title = "{Anatomy of a blazar in the (sub-)millimeter: ALMA observations of PKS 0521-365}",
      journal = {\aap},
         year = 2016,
        month = feb,
       volume = {586},
          eid = {A70},
        pages = {A70},
          doi = {10.1051/0004-6361/201527146},
archivePrefix = {arXiv},
       eprint = {1510.07536},
 primaryClass = {astro-ph.GA},
       adsurl = {https://ui.adsabs.harvard.edu/abs/2016A&A...586A..70L}
}

@ARTICLE{2013MNRAS.434.3487A,
       author = {{An}, Tao and {Baan}, Willem A. and {Wang}, Jun-Yi and {Wang}, Yu and {Hong}, Xiao-Yu},
        title = "{Periodic radio variabilities in NRAO 530: a jet-disc connection?}",
      journal = {\mnras},
         year = 2013,
        month = oct,
       volume = {434},
       number = {4},
        pages = {3487-3496},
          doi = {10.1093/mnras/stt1265},
       adsurl = {https://ui.adsabs.harvard.edu/abs/2013MNRAS.434.3487A}
}

@ARTICLE{1980Natur.287..307B,
       author = {{Begelman}, M.~C. and {Blandford}, R.~D. and {Rees}, M.~J.},
        title = "{Massive black hole binaries in active galactic nuclei}",
      journal = {Nature},
         year = 1980,
        month = sep,
       volume = {287},
       number = {5780},
        pages = {307-309},
          doi = {10.1038/287307a0},
       adsurl = {https://ui.adsabs.harvard.edu/abs/1980Natur.287..307B}
}

@ARTICLE{2021ApJ...919...58Z,
       author = {{Zhang}, Haiyun and {Yan}, Dahai and {Zhang}, Pengfei and {Yang}, Shenbang and {Zhang}, Li},
        title = "{A Quasi-periodic Oscillation in the {\ensuremath{\gamma}}-Ray Emission from the Non-blazar Active Galactic Nucleus PKS 0521-36}",
      journal = {\apj},
         year = 2021,
        month = sep,
       volume = {919},
       number = {1},
          eid = {58},
        pages = {58},
          doi = {10.3847/1538-4357/ac0cf0},
archivePrefix = {arXiv},
       eprint = {2106.10040},
 primaryClass = {astro-ph.HE},
       adsurl = {https://ui.adsabs.harvard.edu/abs/2021ApJ...919...58Z}
}

@ARTICLE{2023arXiv231212623S,
       author = {{Sharma}, Ajay and {Prince}, Raj and {Bose}, Debanjan},
        title = "{Detection of gamma-ray quasi-periodic oscillations in non-blazar AGN PKS 0521-36}",
      journal = {arXiv e-prints},
         year = 2023,
        month = dec,
          eid = {arXiv:2312.12623},
        pages = {arXiv:2312.12623},
          doi = {10.48550/arXiv.2312.12623},
archivePrefix = {arXiv},
       eprint = {2312.12623},
 primaryClass = {astro-ph.HE},
       adsurl = {https://ui.adsabs.harvard.edu/abs/2023arXiv231212623S}
}

@ARTICLE{2017MNRAS.470L.107J,
       author = {{Jim{\'e}nez-Andrade}, E.~F. and {Chavushyan}, V. and {Le{\'o}n-Tavares}, J. and {Pati{\~n}o-{\'A}lvarez}, V.~M. and {Olgu{\'\i}n-Iglesias}, A. and {Kotilainen}, J. and {Falomo}, R. and {Hyv{\"o}nen}, T.},
        title = "{Detection of helicoidal motion in the optical jet of PKS 0521-365}",
      journal = {\mnras},
         year = 2017,
        month = sep,
       volume = {470},
       number = {1},
        pages = {L107-L111},
          doi = {10.1093/mnrasl/slx090},
archivePrefix = {arXiv},
       eprint = {1706.01286},
 primaryClass = {astro-ph.GA},
       adsurl = {https://ui.adsabs.harvard.edu/abs/2017MNRAS.470L.107J}
}

@INPROCEEDINGS{2019ASPC..523..397B,
       author = {{Brasseur}, C.~E. and {Phillip}, Carlita and {Hargis}, Jonathan and {Mullally}, Susan and {Fleming}, Scott and {Fox}, Mike and {Smith}, Arfon},
        title = "{Astrocut: A Cutout Service for TESS Full-Frame Image Sets}",
    booktitle = {Astronomical Data Analysis Software and Systems XXVII},
         year = 2019,
       editor = {{Teuben}, Peter J. and {Pound}, Marc W. and {Thomas}, Brian A. and {Warner}, Elizabeth M.},
       series = {Astronomical Society of the Pacific Conference Series},
       volume = {523},
        month = oct,
        pages = {397},
       adsurl = {https://ui.adsabs.harvard.edu/abs/2019ASPC..523..397B}
}

@software{2018soft12013L,
       author = {{Lightkurve Collaboration} and {Cardoso}, Jos{\'e} Vin{\'\i}cius de Miranda and {Hedges}, Christina and {Gully-Santiago}, Michael and {Saunders}, Nicholas and {Cody}, Ann Marie and {Barclay}, Thomas and {Hall}, Oliver and {Sagear}, Sheila and {Turtelboom}, Emma and {Zhang}, Johnny and {Tzanidakis}, Andy and {Mighell}, Ken and {Coughlin}, Jeff and {Bell}, Keaton and {Berta-Thompson}, Zach and {Williams}, Peter and {Dotson}, Jessie and {Barentsen}, Geert},
        title = "{Lightkurve: Kepler and TESS time series analysis in Python}",
 howpublished = {Astrophysics Source Code Library, record ascl:1812.013},
         year = 2018,
        month = dec,
          eid = {ascl:1812.013},
archivePrefix = {ascl},
       eprint = {1812.013},
       adsurl = {https://ui.adsabs.harvard.edu/abs/2018ascl.soft12013L}
}

@ARTICLE{2026MNRAS.545f1920T,
       author = {{Tripathi}, Ashutosh and {Wiita}, Paul J. and {Dingler}, Ryne and {Smith}, Krista Lynne and {Phillipson}, R.~A. and {Graham}, Matthew J. and {Cui}, Lang},
        title = "{Application of time-series analysis methods to a multiple-sector TESS observations: the case of the radio-loud blazar 3C 371}",
      journal = {\mnras},
         year = 2026,
        month = jan,
       volume = {545},
       number = {2},
          eid = {staf1920},
        pages = {staf1920},
          doi = {10.1093/mnras/staf1920},
archivePrefix = {arXiv},
       eprint = {2512.10533},
 primaryClass = {astro-ph.HE},
       adsurl = {https://ui.adsabs.harvard.edu/abs/2026MNRAS.545f1920T}
}

@article{vaughan2003characterizing,
  title={On characterizing the variability properties of X-ray light curves from active galaxies},
  author={Vaughan, S and Edelson, R and Warwick, RS and Uttley, P},
  journal={Monthly Notices of the Royal Astronomical Society},
  volume={345},
  number={4},
  pages={1271--1284},
  year={2003},
  publisher={Blackwell Science Ltd Oxford, UK}
}

@article{nandra1997asca,
  title={ASCA observations of Seyfert 1 galaxies. I. Data analysis, imaging, and timing},
  author={Nandra, K and George, IM and Mushotzky, RF and Turner, TJ and Yaqoob, T},
  journal={The Astrophysical Journal},
  volume={476},
  number={1},
  pages={70},
  year={1997},
  publisher={IOP Publishing}
}

@article{poutanen2008superorbital,
  title={Superorbital variability of X-ray and radio emission of Cyg X-1--II. Dependence of the orbital modulation and spectral hardness on the superorbital phase},
  author={Poutanen, Juri and Zdziarski, Andrzej A and Ibragimov, Askar},
  journal={Monthly Notices of the Royal Astronomical Society},
  volume={389},
  number={3},
  pages={1427--1438},
  year={2008},
  publisher={Blackwell Publishing Ltd Oxford, UK}
}

@article{edelson2002x,
  title={X-ray spectral variability and rapid variability of the soft X-ray spectrum Seyfert 1 galaxies Arakelian 564 and Ton S180},
  author={Edelson, Rick and Turner, TJ and Pounds, Ken and Vaughan, Simon and Markowitz, Alex and Marshall, Herman and Dobbie, Paul and Warwick, Robert},
  journal={The Astrophysical Journal},
  volume={568},
  number={2},
  pages={610},
  year={2002},
  publisher={IOP Publishing}
}

@ARTICLE{2013PASP..125..306F,
       author = {{Foreman-Mackey}, Daniel and {Hogg}, David W. and {Lang}, Dustin and {Goodman}, Jonathan},
        title = "{emcee: The MCMC Hammer}",
      journal = {\pasp},
         year = 2013,
        month = mar,
       volume = {125},
       number = {925},
        pages = {306},
          doi = {10.1086/670067},
archivePrefix = {arXiv},
       eprint = {1202.3665},
 primaryClass = {astro-ph.IM},
       adsurl = {https://ui.adsabs.harvard.edu/abs/2013PASP..125..306F}
}

@ARTICLE{2012A&A...544A..80G,
       author = {{Gonz{\'a}lez-Mart{\'\i}n}, O. and {Vaughan}, S.},
        title = "{X-ray variability of 104 active galactic nuclei. XMM-Newton power-spectrum density profiles}",
      journal = {\aap},
         year = 2012,
        month = aug,
       volume = {544},
          eid = {A80},
        pages = {A80},
          doi = {10.1051/0004-6361/201219008},
archivePrefix = {arXiv},
       eprint = {1205.4255},
 primaryClass = {astro-ph.HE},
       adsurl = {https://ui.adsabs.harvard.edu/abs/2012A&A...544A..80G}
}

@ARTICLE{1976Ap&SS..39..447L,
       author = {{Lomb}, N.~R.},
        title = "{Least-Squares Frequency Analysis of Unequally Spaced Data}",
      journal = {Astrophysics and Space Science},
         year = 1976,
        month = feb,
       volume = {39},
       number = {2},
        pages = {447-462},
          doi = {10.1007/BF00648343},
       adsurl = {https://ui.adsabs.harvard.edu/abs/1976Ap&SS..39..447L}
}

@ARTICLE{1982ApJ...263..835S,
       author = {{Scargle}, J.~D.},
        title = "{Studies in astronomical time series analysis. II. Statistical aspects of spectral analysis of unevenly spaced data.}",
      journal = {\apj},
         year = 1982,
        month = dec,
       volume = {263},
        pages = {835-853},
          doi = {10.1086/160554},
       adsurl = {https://ui.adsabs.harvard.edu/abs/1982ApJ...263..835S}
}

@ARTICLE{2018ApJS..236...16V,
       author = {{VanderPlas}, Jacob T.},
        title = "{Understanding the Lomb-Scargle Periodogram}",
      journal = {\apjs},
         year = 2018,
        month = may,
       volume = {236},
       number = {1},
          eid = {16},
        pages = {16},
          doi = {10.3847/1538-4365/aab766},
archivePrefix = {arXiv},
       eprint = {1703.09824},
 primaryClass = {astro-ph.IM},
       adsurl = {https://ui.adsabs.harvard.edu/abs/2018ApJS..236...16V}
}

@ARTICLE{2009A&A...496..577Z,
       author = {{Zechmeister}, M. and {K{\"u}rster}, M.},
        title = "{The generalised Lomb-Scargle periodogram. A new formalism for the floating-mean and Keplerian periodograms}",
      journal = {\aap},
         year = 2009,
        month = mar,
       volume = {496},
       number = {2},
        pages = {577-584},
          doi = {10.1051/0004-6361:200811296},
archivePrefix = {arXiv},
       eprint = {0901.2573},
 primaryClass = {astro-ph.IM},
       adsurl = {https://ui.adsabs.harvard.edu/abs/2009A&A...496..577Z}
}

@ARTICLE{2010MNRAS.402..307V,
       author = {{Vaughan}, S.},
        title = "{A Bayesian test for periodic signals in red noise}",
      journal = {\mnras},
         year = 2010,
        month = feb,
       volume = {402},
       number = {1},
        pages = {307-320},
          doi = {10.1111/j.1365-2966.2009.15868.x},
archivePrefix = {arXiv},
       eprint = {0910.2706},
 primaryClass = {astro-ph.HE},
       adsurl = {https://ui.adsabs.harvard.edu/abs/2010MNRAS.402..307V}
}

@ARTICLE{2013MNRAS.433..907E,
       author = {{Emmanoulopoulos}, D. and {McHardy}, I.~M. and {Papadakis}, I.~E.},
        title = "{Generating artificial light curves: revisited and updated}",
      journal = {\mnras},
         year = 2013,
        month = aug,
       volume = {433},
       number = {2},
        pages = {907-927},
          doi = {10.1093/mnras/stt764},
archivePrefix = {arXiv},
       eprint = {1305.0304},
 primaryClass = {astro-ph.IM},
       adsurl = {https://ui.adsabs.harvard.edu/abs/2013MNRAS.433..907E}
}

@ARTICLE{2020NatMe..17..261V,
       author = {{Virtanen}, Pauli and {Gommers}, Ralf and {Oliphant}, Travis E. and {Haberland}, Matt and {Reddy}, Tyler and {Cournapeau}, David and {Burovski}, Evgeni and {Peterson}, Pearu and {Weckesser}, Warren and {Bright}, Jonathan and {van der Walt}, St{\'e}fan J. and {Brett}, Matthew and {Wilson}, Joshua and {Millman}, K. Jarrod and {Mayorov}, Nikolay and {Nelson}, Andrew R.~J. and {Jones}, Eric and {Kern}, Robert and {Larson}, Eric and {Carey}, C.~J. and {Polat}, {\.I}lhan and {Feng}, Yu and {Moore}, Eric W. and {VanderPlas}, Jake and {Laxalde}, Denis and {Perktold}, Josef and {Cimrman}, Robert and {Henriksen}, Ian and {Quintero}, E.~A. and {Harris}, Charles R. and {Archibald}, Anne M. and {Ribeiro}, Ant{\^o}nio H. and {Pedregosa}, Fabian and {van Mulbregt}, Paul and {SciPy 1.  0 Contributors}},
        title = "{SciPy 1.0: fundamental algorithms for scientific computing in Python}",
      journal = {Nature Methods},
         year = 2020,
        month = feb,
       volume = {17},
        pages = {261-272},
          doi = {10.1038/s41592-019-0686-2},
archivePrefix = {arXiv},
       eprint = {1907.10121},
 primaryClass = {cs.MS},
       adsurl = {https://ui.adsabs.harvard.edu/abs/2020NatMe..17..261V}
}

@ARTICLE{2025MNRAS.539.2185M,
       author = {{Malik}, Zahoor and {Akbar}, Sikandar and {Shah}, Zahir and {Misra}, Ranjeev and {Dar}, Athar A. and {Manzoor}, Aaqib and {Ahanger}, Sajad and {Nazir}, Zeeshan and {Iqbal}, Naseer and {Rubab}, Seemin and {Tantry}, Javaid},
        title = "{Statistical Insights into flux and photon index distributions of VHE FSRQs from Fermi-LAT observations}",
      journal = {\mnras},
         year = 2025,
        month = may,
       volume = {539},
       number = {3},
        pages = {2185-2201},
          doi = {10.1093/mnras/staf620},
archivePrefix = {arXiv},
       eprint = {2505.23338},
 primaryClass = {astro-ph.HE},
       adsurl = {https://ui.adsabs.harvard.edu/abs/2025MNRAS.539.2185M}
}

@ARTICLE{2024ApJ...977..111A,
       author = {{Akbar}, Sikandar and {Shah}, Zahir and {Misra}, Ranjeev and {Iqbal}, Naseer},
        title = "{Insights into the Long-term Flaring Events of Blazar PKS 0805-07: A Multiwavelength Analysis Over the Period of 2009{\textendash}2023}",
      journal = {\apj},
         year = 2024,
        month = dec,
       volume = {977},
       number = {1},
          eid = {111},
        pages = {111},
          doi = {10.3847/1538-4357/ad8ddb},
archivePrefix = {arXiv},
       eprint = {2410.23181},
 primaryClass = {astro-ph.HE},
       adsurl = {https://ui.adsabs.harvard.edu/abs/2024ApJ...977..111A}
}

@ARTICLE{2024MNRAS.527.2672S,
       author = {{Sharma}, Ajay and {Kamaram}, Sushanth Reddy and {Prince}, Raj and {Khatoon}, Rukaiya and {Bose}, Debanjan},
        title = "{Probing the disc-jet coupling in S4 0954+65, PKS 0903-57, and 4C +01.02 with {\ensuremath{\gamma}}-rays}",
      journal = {\mnras},
         year = 2024,
        month = jan,
       volume = {527},
       number = {2},
        pages = {2672-2686},
          doi = {10.1093/mnras/stad3399},
archivePrefix = {arXiv},
       eprint = {2311.01738},
 primaryClass = {astro-ph.HE},
       adsurl = {https://ui.adsabs.harvard.edu/abs/2024MNRAS.527.2672S}
}

@article{shah2018log,
  title={Log-normal flux distribution of bright Fermi blazars},
  author={Shah, Zahir and Mankuzhiyil, Nijil and Sinha, Atreyee and Misra, Ranjeev and Sahayanathan, Sunder and Iqbal, Naseer},
  journal={Research in Astronomy and Astrophysics},
  volume={18},
  number={11},
  pages={141},
  year={2018},
  publisher={IOP Publishing}
}

@article{10.1093/mnrasl/sly136,
    author = {Sinha, Atreyee and Khatoon, Rukaiya and Misra, Ranjeev and Sahayanathan, Sunder and Mandal, Soma and Gogoi, Rupjyoti and Bhatt, Nilay},
    title = "{The flux distribution of individual blazars as a key to understand the dynamics of particle acceleration}",
    journal = {Monthly Notices of the Royal Astronomical Society: Letters},
    volume = {480},
    number = {1},
    pages = {L116-L120},
    year = {2018},
    month = {07},
    issn = {1745-3925},
    doi = {10.1093/mnrasl/sly136},
    url = {https://doi.org/10.1093/mnrasl/sly136},
    eprint = {https://academic.oup.com/mnrasl/article-pdf/480/1/L116/25447454/sly136.pdf},
}

@article{10.1093/mnras/stz3108,
    author = {Khatoon, Rukaiya and Shah, Zahir and Misra, Ranjeev and Gogoi, Rupjyoti},
    title = "{Study of long-term flux and photon index distributions of blazars using RXTE observations}",
    journal = {Monthly Notices of the Royal Astronomical Society},
    volume = {491},
    number = {2},
    pages = {1934-1940},
    year = {2019},
    month = {11},
    issn = {0035-8711},
    doi = {10.1093/mnras/stz3108},
    url = {https://doi.org/10.1093/mnras/stz3108},
    eprint = {https://academic.oup.com/mnras/article-pdf/491/2/1934/31160685/stz3108.pdf},
}

@article{Uttley_2001,
    author = {Uttley, Philip and McHardy, Ian M.},
    title = "{The flux-dependent amplitude of broadband noise variability in X-ray binaries and active galaxies}",
    journal = {Monthly Notices of the Royal Astronomical Society},
    volume = {323},
    number = {2},
    pages = {L26-L30},
    year = {2001},
    month = {05},
    issn = {0035-8711},
    doi = {10.1046/j.1365-8711.2001.04496.x},
    url = {https://doi.org/10.1046/j.1365-8711.2001.04496.x},
    eprint = {https://academic.oup.com/mnras/article-pdf/323/2/L26/4079417/323-2-L26.pdf},
}

@ARTICLE{2018Galax...6....2B,
       author = {{Bhatta}, Gopal and {Webb}, James},
        title = "{Microvariability in BL Lacertae: ``Zooming'' into the Innermost Blazar Regions}",
      journal = {Galaxies},
         year = 2018,
        month = jan,
       volume = {6},
       number = {1},
        pages = {2},
          doi = {10.3390/galaxies6010002},
archivePrefix = {arXiv},
       eprint = {1711.08698},
 primaryClass = {astro-ph.HE},
       adsurl = {https://ui.adsabs.harvard.edu/abs/2018Galax...6....2B}
}

@ARTICLE{2026ApJ...998..317K,
       author = {{Kishore}, Shubham and {Gupta}, Alok C. and {Wiita}, Paul J.},
        title = "{Variability Study and Searching for Quasiperiodic Oscillations with Day-like Periods in the Blazar S5 0716+714 with TESS}",
      journal = {\apj},
         year = 2026,
        month = feb,
       volume = {998},
       number = {2},
          eid = {317},
        pages = {317},
          doi = {10.3847/1538-4357/ae3ca0},
archivePrefix = {arXiv},
       eprint = {2603.06099},
 primaryClass = {astro-ph.HE},
       adsurl = {https://ui.adsabs.harvard.edu/abs/2026ApJ...998..317K}
}

@ARTICLE{2026arXiv260303984A,
       author = {{Akbar}, Sikandar and {Shah}, Zahir and {Iqbal}, Naseer},
        title = "{A Short-Timescale Optical Quasi-Periodic Oscillation in PKS\textbackslash,0805$-$07 from High-Cadence TESS Observations}",
      journal = {arXiv e-prints},
         year = 2026,
        month = mar,
          eid = {arXiv:2603.03984},
        pages = {arXiv:2603.03984},
          doi = {10.48550/arXiv.2603.03984},
archivePrefix = {arXiv},
       eprint = {2603.03984},
 primaryClass = {astro-ph.HE},
       adsurl = {https://ui.adsabs.harvard.edu/abs/2026arXiv260303984A}
}

@ARTICLE{2014ApJ...791...21F,
       author = {{Finke}, Justin D. and {Becker}, Peter A.},
        title = "{Fourier Analysis of Blazar Variability}",
      journal = {\apj},
         year = 2014,
        month = aug,
       volume = {791},
       number = {1},
          eid = {21},
        pages = {21},
          doi = {10.1088/0004-637X/791/1/21},
archivePrefix = {arXiv},
       eprint = {1406.2333},
 primaryClass = {astro-ph.GA},
       adsurl = {https://ui.adsabs.harvard.edu/abs/2014ApJ...791...21F}
}





 

\end{document}